\documentclass[useAMS,onecolumn,usenatbib,usegraphicx]{mn2e}

\def\be{\begin{equation}}
\def\ee{\end{equation}}
\def\ba{\begin{eqnarray}}
\def\ea{\end{eqnarray}}
\def\go{\mathrel{\raise.3ex\hbox{$>$}\mkern-14mu\lower0.6ex\hbox{$\sim$}}}
\def\lo{\mathrel{\raise.3ex\hbox{$<$}\mkern-14mu\lower0.6ex\hbox{$\sim$}}}

\begin{document}

\title[Atmosphere Models of Magnetized Neutron Stars] 
{Atmosphere Models of Magnetized Neutron Stars: QED Effects, Radiation Spectra and Polarization Signals}
\author[M. van Adelsberg and D. Lai] {M. van
Adelsberg$^{1}$\thanks{Email: mvanadel@astro.cornell.edu (MVA);
dong@astro.cornell.edu (DL)} and D. Lai$^{1}$\footnotemark[1]\\
$^{1}$Center for Radiophysics and Space Research, Department of
Astronomy, Cornell University, Ithaca, NY 14853}

\date{Accepted 2006 September 21. Received 2006 September 05; 
in original form 2006 July 09}

\pagerange{\pageref{firstpage}--\pageref{lastpage}} \pubyear{2006}

\maketitle
\label{firstpage}
\begin{abstract}
Observations of surface emission from isolated neutron stars (NSs)
provide unique challenges to theoretical modeling of radiative
transfer in magnetized NS atmospheres.  Recent work has demonstrated
the critical role of vacuum polarization effects in determining NS
spectra and polarization signals, in particular the conversion of
photon modes (due to the ``vacuum resonance'' between plasma and vacuum
polarization) propagating in the density gradient of the NS
atmosphere. Previous NS atmosphere models incorporated the mode 
conversion effect approximately, relying on transfer equations
for the photon modes. Such treatments are inadequate near the vacuum 
resonance, particularly for magnetic field strengths around
$B\sim B_l\simeq 7\times 10^{13}$~G, where the vacuum resonance
occurs near the photosphere.
In this paper, we provide an accurate treatment of the mode conversion
effect in magnetized NS atmosphere models, 
employing both the modal radiative transfer equations coupled with 
an accurate mode conversion probability at the vaccum resonance,
and the full evolution equations for the photon Stokes parameters. In doing
so, we are able to quantitatively calculate 
the effects of vacuum polarization on atmosphere structure, emission
spectra and beam patterns, and polarizations for the entire range 
of magnetic field strengths, $B=10^{12}-10^{15}$~G.  
In agreement with previous works, we find that 
for NSs with magnetic field strength $B\ga 2\,B_l$,
vacuum polarization reduces the widths of spectral features, 
and softens the hard spectral tail typical of magnetized atmosphere models.
For $B\la B_l/2$, vacuum polarization does not change the 
emission spectra, but can significantly affect the polarization signals.
Our new, accurate treatment of vacuum polarization 
is particularly important for quantitative 
modeling of NS atmospheres with ``intermediate'' magnetic fields,
$B\simeq (0.5-2)\,B_l$. We provide fitting formulae for the temperature
profiles for a suite of NS atmosphere models with different 
field strengths, effective temperatures and chemical 
compositions (ionized H or He). These analytical profiles are useful for 
direct modeling of various observed properties of NS surface emission.
As an example, we calculate the observed intensity and
polarization lightcurves from a 
rotating NS hotspot, taking into account the evolution of photon 
polarization in the magnetosphere. We show that vacuum polarization
induces a unique energy-dependent linear polarization signature,
and that circular polarization can be generated in the magnetosphere 
of rapidly rotating 
NSs. We discuss the implications of our results to recent 
observations of thermally emitting isolated NSs and magnetars,
as well as the prospects of future spectral and polarization observations.
\end{abstract}



\section{Introduction}
\label{sect:Introduction}

Thermal surface emission from isolated neutron stars
(NSs) \citep[e.g.,][]{Kaspietal05a}
can potentially provide invaluable information on the physical
properties and evolution of NS equations of state at super-nuclear
densities, cooling histories, magnetic fields, and surface
compositions (see, e.g., Prakash et
al.~2001; Yakovlev \& Pethick 2004 for review). 
In recent years, significant progress has been made 
in detecting such radiation using X-ray telescopes 
such as {\it Chandra} and {\it XMM-Newton}.
For example, the spectra of a number of radio pulsars 
(PSR~B1055-52, B0656+14, Geminga and Vela) have been observed to
possess thermal components that can be attributed to emission from NS
surfaces and/or heated polar caps (see, e.g., Becker \& Pavlov
2002). Phase-resolved spectroscopic observations have become 
possible, revealing the surface magnetic field geometry and emission
radius of the pulsar (e.g., Caraveo et al.~2004; De Luca et
al.~2005; Jackson \& Halpern 2005). 
{\it Chandra} has also uncovered a number of compact
sources in supernova remnants with spectra consistent with thermal
emission from NSs (see Pavlov et al.~2003), and useful constraints on
NS cooling physics have been obtained (Slane et al.~2002;
Yakovlev \& Pethick 2004).

Surface X-ray emission has also been detected
from a number of soft gamma-ray repeaters (SGRs) and anomalous X-ray
pulsars (AXPs) --- these are thought to be magnetars, which emit radiation
powered by the decay of superstrong ($B\go 10^{14}$~G) magnetic
fields (see Thompson \& Duncan 1995,1996; Woods \& Thompson 2005).  
The quiescent magnetar emission consists of 
a blackbody-like component with temperature (3--7) $\times 10^6$~K,
and a power-law tail from $2-10$~keV with photon index $2-3.5$
(e.g., Juett et al 2002; Kulkarni et al.~2003; Tiengo et al.~2005),
as well as significant flux at $\sim 100$~keV (Kuiper et al.~2004;
Kuiper et al.~2006). Somewhat surprisingly, the observed 
thermal emission does not show any of the spectral features one might 
expect, such as the ion cyclotron line around 1~keV (for typical magnetar 
field strengths). 

The seven isolated, radio-quiet NSs
(the so-called ``dim isolated NSs''; see Haberl 2005) are also 
of great interest.  These NSs share
the property that their spectra appear to be entirely thermal,
indicating that the emission arises directly from the NS atmospheres,
uncontaminated by magnetospheric processes.  
The true nature of these sources is unclear at present: they could be
young cooling NSs, or NSs kept hot by accretion from the ISM, or
magnetars and their descendants (see van Kerkwijk \& Kulkarni 2001,
Mori \& Ruderman 2003; Haberl 2005; Kaspi et al.~2005). 
While the brightest of these, RX J1856.5-3754, has a featureless
spectrum remarkably well described by a blackbody (Drake et al.~2002;
Burwitz et al.~2003), absorption lines/features at $E\simeq
0.2$--$2$~keV have recently been detected from at least four sources,
including 1E 1207.4-5209 (at 0.7 and 1.4~keV, possibly also 2.1,~2.8~keV;
%
%
Sanwal et al.~2002; Bignami et al.~2003; DeLuca et al.~2004; Mori et al.~2005), 
RX J1308.6+2127 (0.2-0.3~keV; Haberl et al.~2003), RX J1605.3+3249
(0.45~keV; van Kerkwijk et al.~2004), RX J0720.4$-$3125 (0.27~keV;
Haberl et al.~2004a), and possibly several additional sources 
\citep[see][]{Haberletal04b,Zaneetal05a}.
The identification of these features remains
uncertain, with suggestions ranging from electron/ion cyclotron lines
to atomic transitions of H, He or mid-Z atoms in a strong magnetic
field (Ho \& Lai 2004; Pavlov \& Bezchastnov 2005; Mori et al.~2005).
These sources also have different X-ray lightcurves: for example, RX
J1856.5-3754 and RX J1605+3249 show no variability (pulse fraction
$\lo 1$-$3\%$), while RX J0720-3125 shows single-peaked 
pulsations ($P=3.39$~s) of amplitude $\sim 11\%$, with the spectral hardness 
and line width varying with the pulse phase, and RX J1308+2127 shows 
double-peaked pulsations ($P=10.3$~s) with amplitude $\sim 18\%$.
Another puzzle concerns the optical emission: for at least four of these
sources, optical counterparts have been identified, however, the
optical flux is larger (by a factor of $4$-$10$) than the
extrapolation from the blackbody fit to the X-ray spectrum 
%
%
\citep[see, e.g.,][]{Trumperetal04a,Haberl05a}.

The spectrum of NS thermal radiation is formed in the atmosphere layer
(with scale height $\sim 0.1-10$~cm and density $10^{-3}-10^3$~g~cm$^{-3}$)
that covers the stellar surface. Thus, to  
properly interpret observations of NS surface emission, 
detailed modeling of NS atmospheres in strong magnetic fields is required.


\subsection{Previous work on magnetic neutron star atmosphere models}

The first magnetic NS atmosphere models were constructed
by Shibanov et al.~(1992) 
%
%
\citep[see also][]{Pavlovetal95a,Rajagopaletal97a,ZavlinPavlov02a},
who focused on moderate field strengths 
$B\sim 10^{12}$--$10^{13}$~G and assumed full ionization 
%
%
(see also Zane et al.~2000 for atmosphere models with accretion).\footnote{An earlier 
attempt by \citet[][]{Miller92a} adopted a 
polarization-averaging procedure for the radiative transport which is rather
inaccurate since most of the photon flux is carried by the low-opacity photon
mode.}
Similar ionized models for the magnetar field regime ($B\go 10^{14}$~G)
were studied by Zane et al.~(2001), \"Ozel (2001) and Ho \& Lai (2001,2003).
An inaccurate treatment of the free-free opacities in the earlier models
(Pavlov et al.~1995) was corrected by Potekhin \& Chabrier (2003), and
the correction has been incoporated into later models 
(Ho et al.~2003,2004). Recent works (Lai \& Ho 2002,2003a; Ho \& Lai 2003)
have shown that in the magnetar field regime, the effect of
strong-field quantum electrodynamics significantly influences 
the emergent atmosphere spectrum. In particular, vacuum polarization
gives rise to a resonance phenomenon, in which photons can convert 
from the high-opacity mode to the low-opacity one and vice versa.
This vacuum resonance tends to soften the 
hard spectral tail due to the non-greyness of the 
atmosphere and suppress the width of absoprtion lines (see Lai \& Ho 2003a
for a qualitative explanation). Even for modest field strengths 
($B\lo 10^{14}$~G), vacuum polarization can still leave a unique 
%
%
imprint on the X-ray polarization signal (Lai \& Ho 2003b).\footnote{\citet[][]{BulikMiller97a} 
studied the effect of vacuum polarization on Compton scattering in an isothermal magnetized 
plasma, in the context of SGR bursts.  They did not calculate self-consistent atmosphere 
models and did not find the effects (suppression of spectral lines and 
hard tails) discussed in the works of Ho \& Lai. \citet[][]{Zaneetal01a} included vacuum 
polarization effects in their models, but did not report any of the important effects of QED. 
\citet[][]{Ozel01a} also included vacuum polarization effects, but came to 
the opposite conclusion (i.e., that vacuum polarization effects cause thermal spectral 
tails to become harder), which is incorrect.}

Because a strong magnetic field greatly increases the binding 
energies of atoms, molecules and other bound species (see Lai 2001),
these bound states may have appreciable abundances in the NS atmosphere
(Lai \& Salpeter 1997; Potekhin et al.~1999). 
%
%
Early considerations of partially ionized atmospheres 
\citep[e.g.,][]{Rajagopaletal97a}
relied on oversimplified treatments of atomic physics (e.g., ionization 
equilibrium and equation of state) and
nonideal plasma effects in strong magnetic fields.  
Recently, a thermodynamically consistent equation of state and opacities 
for magnetized ($B=10^{12}-10^{15}$~G), partially ionized H plasma have been
obtained (Potekhin et al.~1999; Potekhin \& Chabrier 2003,2004),
and the effect of bound atoms on the dielectric tensor of the plasma
has also be studied (Potekhin et al.~2004). These improvements 
have been incorporated 
%
%
into partially ionized, magnetic NS atmosphere models 
(Ho et al.~2003; Potekhin et al.~2004). Finally, for sufficiently 
low temperatures and/or strong magnetic fields, the NS atmosphere
may undergo a phase transition into a condensed state 
%
%
\citep[see][]{Ruderman71a,Jones86a,Neuhauseretal87a,LaiSalpeter97a,Lai01a,
MedinLai06a,MedinLai06b}.
%
Thermal emission from such a condensed surface has been studied by van
Adelsberg et al.~(2005) (cf. Turolla et al.~2004; Perez-Azorin et
al.~2005).


\subsection{This paper}

While previous works have identified 
the importance and trend of the vacuum polarization effect
(Lai \& Ho 2002,2003a), the implementation of
the effect in NS atmosphere models (Ho \& Lai 2003,2004)
has been based on approximations (see \S 2).
So far all studies of magnetic
NS atmospheres have relied on solving the transfer equations for the
specific intensities of the two photon modes.
As discussed in Lai \& Ho (2003a) and reviewed in \S 2 below, 
these equations cannot properly handle the vacuum-induced mode conversion
phenomenon because mode conversion intrinsically involves 
interference between different modes.

In this paper we provide a new, quantitatively accurate 
treatment of vacuum polarization effects in radiation transfer for 
fully ionized NS atmospheres. Our work confirms
the semi-quantitative results obtained in 
previous works based on an approximate treatment of vacuum resonance
(Ho \& Lai 2003; Lai \& Ho 2003b). Moreover, our new 
treatment allows us to quantitatively predict the spectral and
polarization properties of NS atmospheres with field strengths varying
from $10^{12}$~G to $10^{15}$~G.
 
The remainder of our paper is organized as follows: \S\ref{sect:VacEffect}
describes the basic physics of vacuum polarization in radiative
transfer and resonant mode conversion; \S\ref{sect:Method} details two
methods of solving the radiative transfer problem and our basic
physics inputs, as well as tests of our numerical method;
\S\ref{sect:Results} presents the basic results from our models, including 
atmosphere structures (with fitting formulae for the temperature profiles), 
spectra, and emission beam patterns;
\S\ref{sect:LCurvesPolar} considers the observed 
polarization signals of atmosphere emission for a simple geometry 
(i.e., a rotating NS hotspot); and
\S\ref{sect:Discussion} discusses the implications of our results.


\section{Effect of Vacuum Polarization on Radiative Transfer}
\label{sect:VacEffect}

Before describing our quantitative treatment of the
vacuum polarization effect in NS atmospheres, it is
useful to summarize the basic physics of the effect
(see also Lai \& Ho 2003a) and discuss the limitations
of previous treatments.

\begin{figure}
\includegraphics[width=84mm]{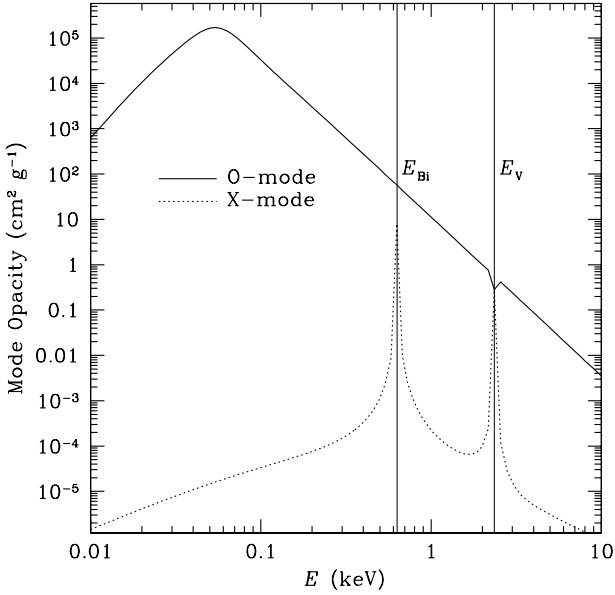}
\caption{Photon free-free absorption opacities for X and O
polarization modes as a function of energy at $B=10^{14}$ G,
$T=10^6$~K, $\theta_{kB}=\pi/4$ and $\rho=5.4$~g~cm$^{-3}$.  
Vacuum polarization induces the sharp resonance feature for the X-mode opacity
at $E_V$. This ``spiky opacity'' can affect the emergent radiation
spectrum from magnetized NSs, but does not include 
all the effects associated with the vacuum resonance.}
\label{fig:Kappa}
\end{figure}

Photons 
(with energy $E\ll E_{Be}=\hbar e B / m_e c$, the electron cyclotron energy)
in magnetized NS atmospheres usually propagate in two distinct 
polarization states, the ordinary mode (denoted by ``O'') 
and the extraordinary mode (denoted by ``X''), which are polarized (almost)
parallel and perpendicular to the plane made by the magnetic field 
and direction of photon propagation, respectively. 
In strong magnetic fields, the dielectric tensor describing the atmospheric 
plasma of a NS must be corrected for QED vacuum effects 
(Gnedin et al.~1978; M\'{e}sz\'{a}ros \& Ventura 1979; Pavlov \& Shibanov 1979;
M\'{e}sz\'{a}ros 1992). For a photon propagating 
in a medium of {\it constant} density $\rho$, the plasma and vacuum contributions 
to the dielectric tensor ``cancel'' each other out at a particular 
energy given by 
\begin{eqnarray}
E_V= 1.02\left({Y_e\,\rho_1}\right)^{1/2}
B_{14}^{-1} f_B\ \mbox{keV},
\label{eq:E_V}
\end{eqnarray}
where $Y_e=Z/A$ ($Z,~A$ are the atomic number and mass number, respectively),
$\rho_1 = \rho/(1\ \rm{g\ cm}^{-3})$, $B_{14}=B/(10^{14}\ \mbox{G})$,
and $f_B\sim 1$ is a slowly varying function of $B$
[see eq.~(2.41) of Ho \& Lai (2003)]. At the resonance, both modes
become circularly polarized. A number of 
previous papers (e.g., M\'{e}sz\'{a}ros 1992) emphasized the 
sharp X-mode opacity feature associated with the resonance 
(see Fig.~\ref{fig:Kappa}). It might seem that to understand the
vacuum polarization effect in radiative transfer, all one needs
to do is to include this spike in the opacity
(e.g., \"Ozel 2003). However, this is not the whole story.


A more useful way to understand the effects of the
vacuum resonance is to consider a photon with given energy $E$,
traversing the density gradient of a NS atmosphere.  The photon will
encounter the vacuum resonance at the density
\begin{eqnarray}
\rho_V = 0.96\, Y_e^{-1} E_1^2\,
B_{14}^2\, f_B^{-2}\ \mbox{g cm}^{-3},
\end{eqnarray}
where $E_1=E/(1\ \mbox{keV})$.
\citet[][]{LaiHo02a} showed that the photon undergoes resonant mode 
conversion when the adiabatic condition $E\ga E_{ad}$ is satisfied, with\footnote{Since $E_{ad}$ depends on
$E$, one needs to solve for $E\ga E_{ad}$ to determine the adiabatic region.  See Fig.~6 of 
\citet[][]{LaiHo03a}.}
\begin{eqnarray}
E_{ad}=
2.52\left[f_B\tan\theta_{kB}\left|1-\left({E_{Bi}/E}\right)^2\right|
\right]^{2/3}\left({1\ \mbox{cm}\over H_{\rho}}\right)^{1/3},
\end{eqnarray}
where $\theta_{kB}$ is the angle between the magnetic field and
direction of propagation, $E_{Bi}=0.63(Z/A)$~keV is the ion cyclotron
energy, and $H_{\rho}\equiv |ds/d\ln\rho|$ is the density scale height
along the ray.  Thus, an adiabatic O-mode photon encountering the
vacuum resonance will convert into an X-mode photon, and vice-versa.
In general, for intermediate energies $E\sim E_{ad}$, photons undergo partial
conversion, in which an O-mode converts to a X-mode (and
vice-versa) with probability $1-P_{\rm jump}$, where $P_{\rm jump}$ is the non-adiabatic 
jump probability
\be
\label{eq:P_j}
P_{\rm jump}=\exp\left[-{\pi\over 2}(E/E_{ad})^3\right].
\ee
Due to free-free absorption, the X-mode opacity is
suppressed relative to the O-mode by a factor of
$(E_{Be}/E)^2$, where the electron cyclotron energy is $E_{Be}=1158~B_{14}$ keV; 
thus, the 
mixing of photon modes at the resonance can have a drastic effect on
the radiative transfer. For magnetic field strengths satisfying
(Lai \& Ho 2003a, Ho \& Lai 2004)
\be
B \go B_l \simeq 6.6\times 10^{13}\, T_6^{-1/8}E_1^{-1/4}S^{-1/4}\mbox{ G},
\label{eq:maglim}
\ee
where $T_6=T/(10^6~{\rm K})$ and
$S=1-e^{-E/kT}$, 
the vacuum resonance density lies between the X-mode
and O-mode photospheres for typical photon energies, 
leading to suppression of spectral features and
softening of the hard X-ray tail characteristic of ionized hydrogen
atmospheres. For ``normal'' magnetic fields, $B\lo B_l$, the vacuum resonance 
lies outside both photospheres, and the emission spectrum is 
unaltered by the vacuum resonance, although the observed polarization 
signals are still affected (Lai \& Ho 2003b).

In their implementation of the vacuum resonance effect in NS
atmosphere models, Ho \& Lai (2003) considered two limiting cases: (i)
complete mode conversion ($P_{\rm jump}=0$), which is equivalent to assuming that 
$E\gg
E_{ad}$ is satisfied for all photon energies; (ii) no conversion ($P_{\rm jump}=1$),
which is equivalent to assuming $E\ll E_{ad}$ for all photons. In
the former case, all X-mode photons are converted to the O-mode at the
resonance (and vice-versa), whereas in the latter, such conversion is
neglected. In both cases, radiative transfer equations based on 
photon modes can be used, as long as one properly defines the modes across the
resonance (Ho \& Lai 2003). We expect that the complete and 
no conversion limits bracket the correct solution.
In case (ii), one only has the
``narrow spiky opacity'' effect associated with the resonance.  Lai \&
Ho (2002) estimated the width of this opacity spike and emphasized the
importance of resolving the spike. In both limits, vacuum resonance
has qualitatively the same effects on the emergent spectrum, i.e.,
suppression of lines and softening of hard spectral tails (Lai \& Ho
2002; Ho \& Lai 2003), although for $B\sim ({\rm a~few}\times
10^{13})-10^{14}$~G, appreciable quantitative differences in the 
spectra using the two limits are produced (Ho \& Lai 2004).

As mentioned before, all studies of radiative transfer in magnetized 
NS atmospheres so far have relied on solving the transfer equations for the
specific intensities of the two photon modes (e.g.,
M\'{e}sz\'{a}ros 1992; Zavlin \& Pavlov 2002). These equations cannot properly 
handle the vacuum-induced mode conversion phenomenon. 
In particular, photons with energies 0.3-2~keV are only partially converted across
the vacuum resonance (this is the energy range
in which the bulk of the radiation emerges and spectral lines are
expected for $B\sim 10^{14}$~G). 
In addition, the phenomenon of mode collapse 
(when the X and O-modes become degenerate) occurs when 
dissipative effects are included in the plasma dielectric tensor,
and the concomitant breakdown of the Faraday
depolarization condition near the resonance further complicates the
standard treatment of radiative transfer based on normal modes.
As shown by \citet[][]{GnedinPavlov74a}, the modal description of radiative 
transfer is valid only in the limit 
$|\mbox{Re}(n_{\rm X}-n_{\rm O})|\gg |\mbox{Im}(n_{\rm X}+n_{\rm O})|$, 
where $n_{\rm X}$ and $n_{\rm O}$
are the indices of refraction corresponding to the X and O-modes,
respectively.  \citet[][]{HoLai03a} showed that, for a narrow range of
energies around the vacuum resonance, this condition can be violated,
and the violation becomes especially pronounced in the magnetar field regime.
It is not obvious whether the mode collapse significantly alters the 
radiative transfer. Thus, to account for the vacuum resonance effect 
in a quantitative manner, one must 
solve the transfer equations in terms of the photon intensity matrix 
(Lai \& Ho 2003a) and properly take into account the probability of
mode conversion. This is one of the main goals of our paper.


\section{Method}
\label{sect:Method}

\subsection{Partial mode conversion using photon mode equations}
\label{subsect:Modes}

\subsubsection{Radiative transfer equation}
\label{subsubsect:PModeRTE}

For our models, we consider plane-parallel, fully ionized H or He atmospheres, with the 
magnetic field oriented normal to the surface.  The standard method used in all previous work
involves solving the coupled radiative transfer equations for the two modes of photon propagation.  
These are given by
\begin{eqnarray}
\label{eq:ModeRTE}
\pm\mu{\partial I^j_{\nu}(\tau,\pm\mu)\over \partial\tau}={\kappa^{tot}_j\over \kappa_T}
\left[I^j_{\nu}(\tau,\pm\mu)-S^j_{\nu}(\tau,\pm\mu)\right]
\end{eqnarray}
where $I^j_{\nu}(\tau,\mu)$ is the specific intensity for mode $j$, 
$\mu={\bf\hat{k}}\cdot{\bf\hat{z}}\ge 0$,
$\kappa^{tot}_j=\kappa_j^{\rm ff}+\kappa_j^{\rm sc}$ is the 
total opacity (with contributions from free-free absorption and scattering, see below), 
$\kappa_T=0.4$ cm$^2$ g$^{-1}$ is the Thomson 
scattering opacity, $\tau$ is the Thomson optical depth (defined by $d\tau=-\rho\,\kappa_T\,dz$), 
and $S^j_{\nu}$ is the source function, defined below.  Eq.~(\ref{eq:ModeRTE}) is solved subject to the 
constraints of hydrostatic and radiative equilibria, as well as constant radiative flux $F_{\rm rad}$, given by:
\begin{eqnarray}
\label{eq:HydroEq}
P = {g\over\kappa_T}\tau,
\end{eqnarray}
\begin{eqnarray}
\label{eq:RadEq1}
\int_0^{\infty} d\nu\sum_{j=1}^2\kappa^{abs}_j\left({B_{\nu}\over 2}-J^j_{\nu}\right)=0,
\end{eqnarray}
\begin{eqnarray}
\label{eq:CFlux}
F_{\rm rad} = 2\pi\sum_{j=1}^2\int_0^{\infty}d\nu\int_0^1d\mu \mu\left[I_{\nu}^j(\mu)-I_{\nu}^j(-\mu)\right]
=\sigma_{\rm sb} T_{\rm eff}^4
\end{eqnarray}
where $P$ is the pressure of electrons and ions, $g = \left({GM\over R^2}\right)\left(1-{2GM\over Rc^2}\right)^{-1/2} 
= 2.4\times 10^{14}$ cm s$^{-2}$ is the 
surface gravitational acceleration (we adopt $M=1.4$ M$_{\sun}$ and $R=10$ km throughout the paper), 
$J^j_{\nu}\equiv (1/2)\int_0^1 d\mu \left[I^j_{\nu}(\mu)+I^j_{\nu}(-\mu)\right]$ is the mean specific intensity, 
$B_{\nu}$ is the Planck function, 
and $T_{\rm eff}$ is the effective temperature of the atmosphere.
To integrate eq.~(\ref{eq:ModeRTE}) subject to 
the conditions of (\ref{eq:HydroEq})-(\ref{eq:CFlux}), we assume the ideal gas equation of state for both 
protons and electrons; electron degeneracy effects are neglected.  \citet[][]{HoLai01a} showed that 
the effect of this approximation on the atmosphere is negligible.
Note that in general, thermal conduction due to electrons also contributes to the total flux.  However, 
we have found that in the atmosphere region of interest, the conduction flux is always less than a 
few percent of the total flux, and is therefore neglected.

\subsubsection{Photon modes and opacities}
\label{subsubsect:MOSF}

The properties of magnetized atmospheric plasma can be described by a complex dielectric tensor 
\citep[][]{Ginzburg64a}.  In a coordinate system with the magnetic field aligned with the $z$-axis, 
the plasma contribution to the dielectric tensor takes the form \citep[][]{LaiHo03a}:\footnote{Note that 
eq.~(13) of 
\citet[][]{LaiHo03a} should be 
$\gamma_{ei}^{\pm}=\gamma_{ei}^{\perp}(1+{m_e\over A m_p}) \approx \gamma_{ei}^{\perp}$.  This 
substitution should be applied to all the appropriate equations in \citet[][]{LaiHo03a}.}
\begin{eqnarray}
\label{eq:DieTen}
\left[\mbox{\boldmath{$\epsilon$}}^{(pl)}\right] = 
\left (
\begin{array}{ccc}
\epsilon & i\,g & 0 \\ 
-i\,g & \epsilon & 0 \\
0 & 0 & \eta
\end{array}
\right ),
\end{eqnarray}
where
\begin{eqnarray}
\label{eq:epg}
\epsilon\pm g & \approx & 1 - {v_e (1 + i\gamma_{ri}) + v_i (1 + i\gamma_{re})\over (1 + i\gamma_{re} \pm u_e^{1/2} )
	(1 + i\gamma_{ri} \mp u_i^{1/2} ) + i\gamma_{ei}^{\perp})}\\
\label{eq:eta}
\eta & \approx & 1 - {v_e \over 1 + i(\gamma_{ei}^{\parallel}+\gamma_{re})} - {v_i \over 1 + i(\gamma_{ei}^{\parallel}+\gamma_{ri})}.
\end{eqnarray}
In eqs.~(\ref{eq:epg})--(\ref{eq:eta}) we have defined the dimensionless ratios $u_e\equiv (E_{Be}/E)^2$, 
$u_i\equiv (E_{Bi}/E)^2$, $v_e\equiv (E_{pe}/E)^2$, $v_i\equiv (E_{pi}/E)^2$, 
where $E_{pe} = \hbar (4\pi n_e e^2 / m_e)^{1/2} = 0.02871 (Y_e\,\rho_1)^{1/2}$ keV is the electron plasma energy, and 
$E_{pi} = (Z m_e / A m_p) E_{pe} = 6.70 \times 10^{-4}\, Y_e\, \rho_1^{1/2}$ keV is the ion plasma energy.
The dimensionless damping rates $\gamma_{ei}^{\perp,\parallel} = \nu_{ei}^{\perp,\parallel}/\omega$
(for electron-ion collisional damping), 
$\gamma_{re}=\nu_{re}/\omega$ (for electron radiative damping), and 
$\gamma_{ri}=\nu_{ri}/\omega$ (for ion radiative damping)
are given by,
\begin{eqnarray}
\gamma_{ei}^{\perp,\parallel} & = & 9.2\times 10^{-5}{Z^2\, \rho_1\over A\, T_6^{1/2}\, E_1^2}\,\left(1-e^{-E/k_B T}\right)\, 
	g^{\rm ff}_{\perp,\parallel},\\
\gamma_{re} & = & 9.5\times 10^{-6} E_1,\\
\gamma_{ri} & = & 5.2\times 10^{-9} {Z^2 \over A} E_1.
\end{eqnarray}
The quantities 
$g^{\rm ff}_{\perp}$ and $g^{\rm ff}_{\parallel}$ are the velocity-averaged magnetic Gaunt factors perpendicular and parallel to 
the magnetic field, respectively; they are calculated using eqs.~(4.4.9)-(4.4.12) 
from \citet[][]{Meszaros92a}.\footnote{Note that eq.~(4.4.12) of \citet[][]{Meszaros92a} should be 
$a_{\pm}=(p\pm[p^2+2m\hbar\omega]^{1/2})^2(2mk_B T)^{-1}$.}
This calculation includes contributions from electrons in the ground Landau level only.  Gaunt factors 
including contributions from excited states have been derived by \citet[][]{PotekhinChabrier04a}.  Nevertheless, for 
energies well below $E_{Be}$, the differences between the two calculations are negligible \citep[][]{Potekhin06a}.

Vacuum contributions to the dielectric tensor can be taken into account by making the following substitutions into 
the tensor of eq.~(\ref{eq:DieTen}):
\begin{eqnarray}
\epsilon\rightarrow\epsilon' = \epsilon + a - 1,\ \ \eta\rightarrow\eta' = \eta + a + q - 1,
\end{eqnarray}
where $a$ and $q$ are vacuum parameters given by the expressions in,
e.g., Heyl \& Hernquist (1997) and Potekhin et al.~(2004) 
(the latter also contains general fitting formulae).
Solving Maxwell's 
equations for the anisotropic medium yields two modes of propagation.  In a coordinate system where the wave vector ${\bf k}$ is 
along the $z$-axis and the magnetic field lies in the 
$xz$ plane (such that $\hat{\bf k}\times\hat{\bf B}=-\sin\theta_{kB}\hat{\bf y}$), the mode eigenvectors can be written as
\begin{eqnarray}
\label{eq:ModeEigenvs}
{\bf e}_{\pm} = {1\over (1+|K_{\pm}|^2+|K_{z\pm}|^2)^{1/2}}(iK_{\pm},1,iK_{z\pm}),
\end{eqnarray}
where the ellipticity $K_{\pm} = -i e_x/e_y$ of mode $\pm$ is given by
\begin{eqnarray}
\label{eq:Ellipticity}
K_{\pm} = \beta\pm\sqrt{\beta^2+r},
\end{eqnarray}
with $r=1+(m/a)\sin^2\theta_{kB}$ ($m$ is another vacuum polarization
parameter; Heyl \& Hernquist 1997; Potekhin et al.~2004),
and the polarization parameter $\beta$ is
\begin{eqnarray}
\beta = -{\epsilon'^2-g^2-\epsilon'\eta'(1+m/a) \over 2 g\eta'}{\sin^2\theta_{kB}\over\cos\theta_{kB}}.
\end{eqnarray}
The $z$-components of the mode eigenvectors are given by
\begin{eqnarray}
K_{z\pm}=-{(\epsilon-\eta-g)\sin\theta_{kB}\cos\theta_{kB} K_{\rm\pm}+g\sin\theta_{kB}\over 
\epsilon\sin^2\theta_{kB}+(\eta+q)\cos^2\theta_{kB}+a-1}.
\end{eqnarray}

Note that when the modes are labelled according to eq.~(\ref{eq:Ellipticity}), the $K_{\pm}$ vary continuously across 
the vacuum resonance 
($\beta=0$), and do not cross each other in the absence of dissipation \citep[][]{LaiHo03a}.  Another 
way of labeling the modes, commonly adopted in the literature \citep[e.g.][]{Meszaros92a}, is 
\begin{eqnarray}
\label{eq:EllipticityXO}
K_j = \beta\left[1+(-1)^j\left(1+{r\over\beta^2}\right)^{1/2}\right].
\end{eqnarray}
According to this labeling scheme, $j=1$ corresponds to the X-mode ($|K_1|<1$) and $j=2$ corresponds to the O-mode ($|K_2|>1$).
Obviously, $K_{1}$ and $K_2$ are not continuous functions across the vacuum resonance.  
It is also clear that a given + mode (or - mode) 
which manifests as the X-mode (O-mode) before the resonance switches character after the resonance.

Using the mode eigenvectors and the componenets of the dielectric tensor, expressions for the free-free absorption and 
scattering opacities can be 
obtained.  The cyclic components of the mode eigenvectors in a rotating frame with the magnetic field along the 
$z$-axis are:
\begin{eqnarray}
\left |e_{\pm}^j\right |^2 & = & \left|{1\over\sqrt{2}}(e_X^j+i e_Y^j)\right|^2 = 
{1\pm|K_j\cos\theta_{kB}+K_{zj}\sin\theta_{kB}|^2\over 2(1+|K_j|^2+|K_{zj}|^2)},\\
\left|e_{o}^j\right|^2 & = & {\left|K_j\sin\theta_{kB}-K_{zj}\cos\theta_{kB}\right|^2\over 1+|K_j|^2+|K_{zj}|^2}.
\end{eqnarray}
Note that in the above expression, $j$ indicates the mode, and the $\pm$ subscript should not be confused with the $K_{\pm}$ 
labeling of photon modes.  The free-free absorption opacity for mode $j$ can be written \citep[][]{LaiHo03a}:
\begin{eqnarray}
\kappa^{\rm ff}_j & = & \kappa_+^j|e_+^j|^2+\kappa_-|e_-^j|^2+\kappa_o|e_o^j|^2,
\end{eqnarray}
with
\begin{eqnarray}
\kappa_{\pm} & = & {\omega\over c\rho}  v_e \Lambda_{\pm}\gamma_{ei}^{\perp},\\
\kappa_o & = & {\omega\over c\rho} v_e\gamma_{ei}^{\parallel},\\
\Lambda_{\pm} & = & \left[(1\pm u_e^{1/2})^2(1\mp u_i^{1/2})^2+\gamma_{\pm}^2\right]^{-1},\\
\gamma_{\pm} & = & \gamma_{ei}^{\perp} + (1\pm u_e^{1/2})\gamma_{ri} + (1\mp u_i^{1/2})\gamma_{re}.
\end{eqnarray}
Note that these expressions include the contribution of electron-ion Coulomb collisions to the free-free absorption 
opacity in a consistent way.  They correct the free-free opacity adopted in earlier papers 
\citep[e.g.][]{Pavlovetal95a,HoLai01a},
and they agree with the correct expressions given in \citet[][]{PotekhinChabrier03a}, and those used by 
\citet[][]{Hoetal03a,Hoetal04a}.

The scattering opacity from 
mode $j$ into mode $i$ is given by \citet[][]{Ventura79a} \citep[see][]{HoLai01a}:
\begin{eqnarray}
\kappa^{sc}_{ji}=Y_e\kappa_T\sum_{\alpha=-1}^1\left[(1+\alpha u_e^{1/2})^2+\gamma_e^2\right]^{-1}
|e_{\alpha}^j|^2 A_{\alpha}^i+\left({Z^2 m_e\over A m_p}\right)^2 {\kappa_T\over A}\sum_{\alpha=-1}^1
\left[(1-\alpha u_i^{1/2})^2+\gamma_i^2\right]^{-1}|e_{\alpha}^j|^2 A_{\alpha}^i
\end{eqnarray}
where $\gamma_e=\gamma_{ei}^{\alpha}+\gamma_{re}$, $\gamma_i=\gamma_{ei}^{\alpha}+\gamma_{ri}$, and 
$A_{\alpha}^i=(3/4)\int_{-1}^1 d\mu' |e_{\alpha}^i|^2$.  The total 
scattering opacity from mode $j$ is then $\kappa_j^{sc}=\sum_i\kappa^{sc}_{ji}$.

\subsubsection{Source function}
\label{subsubsect:SourceFunc}

The source function in eq.~(\ref{eq:ModeRTE}) can be written as
\begin{eqnarray}
S_{\nu}^j(\mu) = {\kappa^{\rm ff}_j(\mu)\over\kappa^{\rm tot}(\mu)}{B_{\nu}\over 2}+
	{2\pi\over\kappa_j^{tot}(\mu)}\sum_{i=1}^2\int_0^1 d\mu' 
	{d\kappa^{sc}(\mu' i\rightarrow \mu j)\over d\Omega}\left[I_{\nu}^j(\mu)+I_{\nu}^i(-\mu)\right]
\end{eqnarray}
where \citep[][]{Ventura79a}
\begin{eqnarray}
{d\kappa^{sc}(j \mu\rightarrow i \mu')\over d\Omega'} = {3\over 8\pi}Y_e \kappa_T\left|\sum_{\alpha=-1}^1 
{1\over 1+\alpha u_e^{1/2}}\, {e^j_{\alpha}}^* e^i_{\alpha}\right|^2 + 
{3\over 8\pi}\left({Z^2 m_e\over A m_p}\right)^2 {\kappa_T\over A} 
\left|\sum_{\alpha=-1}^1 {1\over 1-\alpha u_i^{1/2}}\, {e_{\alpha}^j}^* e_{\alpha}^i\right|^2.
\end{eqnarray}
Following \citet[][]{HoLai01a}, 
it is a good approximation to assume that the differential scattering cross-section is independent of the initial 
photon direction.
The resulting approximate source function is:
\begin{eqnarray}
S^j_{\nu}(\mu)\approx {\kappa_j^{\rm ff}(\mu)\over\kappa_j^{tot}(\mu)}{B_{\nu}\over 2}+\sum_i{\kappa_{ji}^{sc}(\mu)\over 
\kappa_j^{tot}(\mu)}{c u_{\nu}^i\over 4\pi}
\end{eqnarray}
where $u_{\nu}^j=(2\pi/c)\int_{-1}^1 d\mu' I^j_{\nu}(\mu')$ is the specific energy density of mode $j$.
Note that 
the source function depends on the specific intensity in all directions, and thus depends on 
the solution to the radiative transfer equation.  Therefore, we calculate $S_{\nu}^j$ iteratively, 
according to the scheme described in \S\ref{subsubsect:ModeSol}-\S\ref{subsubsect:TempCorrect}.

\subsubsection{Solution to transfer equation for photon modes including partial mode conversion}
\label{subsubsect:ModeSol}

We describe above how vacuum polarization effects can be incorporated into the 
free-free absorption and scattering opacities for the photon modes.  
However, these opacity effects do not capture the essence of the vacuum resonance phenomena.  As discussed in 
\citet[][]{LaiHo03a}, solving the transfer eq.~(\ref{eq:ModeRTE}) 
using $K_{\pm}$ [eq.~(\ref{eq:Ellipticity})] as the 
basis for the photon modes amounts to assuming complete mode conversion ($P_{\rm jump}=0$), 
while using $K_{1,2}$ [eq.~(\ref{eq:EllipticityXO})] 
corresponds to assuming no mode conversion ($P_{\rm jump}=1$).  This was the strategy adopted by previous works.
To correctly account for the vacuum resonance effect, it is necessary to use the jump probability $P_{\rm jump}$
[eq.~(\ref{eq:P_j})]
to convert the mode intensities across the resonance according to the formulae:
\begin{eqnarray}
\label{eq:ModeConv2}
I_X & \rightarrow & P_{\rm jump}\, I_X + (1-P_{\rm jump}) I_O,\\
\label{eq:ModeConv3}
I_O & \rightarrow & P_{\rm jump}\, I_O + (1-P_{\rm jump}) I_X.
\end{eqnarray}
Note that since the resonance density depends on photon energy, the standard Feautrier procedure for integrating the radiative 
transfer equation cannot be used here, as there is no simple way 
to incoporate eqs.~(\ref{eq:ModeConv2})-(\ref{eq:ModeConv3}) into the 
method of forward and 
backward substitution employed by Feautrier \citep[see][\S 6-3]{Mihalas78a}.
Instead, we use the standard Runga-Kutta method to integrate the transfer 
eq.~(\ref{eq:ModeRTE}) in the upward and downward directions starting from the boundary 
conditions:
\begin{eqnarray}
I_{\nu}^j(\tau\rightarrow\tau_{max},+\mu) & \rightarrow & B_{\nu}/2\\
I_{\nu}^j(\tau\rightarrow\tau_{min},-\mu) & \rightarrow & 0
\end{eqnarray}
The Runga-Kutta integration is stopped at the resonance, where eqs.~(\ref{eq:ModeConv2}) and 
(\ref{eq:ModeConv3}) are used to convert the mode intensities.  Then the integration is continued 
to completion.
The limits $\tau_{max}$ and $\tau_{min}$ are chosen to span 5-8 orders of magnitude.  This insures 
that (1)
photons begin their evolution at densities greater than the X-mode decoupling depth, 
and (2) both 
X-mode and O-mode photons are fully decoupled from the matter at the outermost edge of the atmosphere ($\tau=\tau_{min}$).
We finite-difference eq.~(\ref{eq:ModeRTE}) as:
\begin{eqnarray}
\pm\mu {I'-I\over\Delta\tau}\approx {1\over 2\kappa_T}\left[\kappa(I-S)+\kappa'(I'-S')\right]
\end{eqnarray}
where $I=I_{\nu}^j(\tau,\pm\mu)$, $I'=I_{\nu}^j(\tau',\pm\mu)$, $\kappa=\kappa_j^{\rm tot}$, 
$\kappa'=\kappa_j^{\rm tot}(\tau')$, $S=S_{\nu}^j(\tau,\pm\mu)$, $S'=S_{\nu}^j(\tau',\pm\mu)$,
with $\tau'=\tau+\Delta\tau$.  This gives
\begin{eqnarray}
\label{eq:FinDiff}
I'={1\over 1\mp{\Delta\tau\kappa'\over 2\mu\kappa_T}}\left[\left(1\pm{\Delta\tau\kappa\over 2\mu\kappa_T}\right)I\mp 
{\Delta\tau \over 2\mu\kappa_T}(\kappa S+\kappa'S')\right]
\end{eqnarray}
This formula yields stable integrations whose results are not strongly dependent on grid spacing 
(see \S\ref{subsect:GridTests}).

To summarize, our method for integrating the radiative transfer with partial mode conversion 
is as follows: (1) For given $E$ and 
$\theta_{kB}$, we integrate eq.~(\ref{eq:ModeRTE}) using (\ref{eq:FinDiff}) from $\tau_{max}$
to the vacuum resonance at optical depth $\tau_V$ (defined by $\rho(\tau_V)=\rho_V$).
(2) At the resonance, the X-mode and O-mode intensities are converted using eqs.~(\ref{eq:ModeConv2}) and 
(\ref{eq:ModeConv3}).
(3) Integration of eq.~(\ref{eq:ModeRTE}) is continued to $\tau_{min}$.  
We use an analagous procedure for downward integration from
$\tau_{min}$ to $\tau_{max}$. 

\subsubsection{Temperature correction procedure}
\label{subsubsect:TempCorrect}

To integrate the radiative transfer equation, an initial temperature profile is assumed 
(the initial source function is set to $B_{\nu}/2$).  This initial profile is taken 
from a previously constructed model with the same magnetic field and effective temperature, but 
without partial mode conversion \citep[see][]{HoLai03a}.  In general, the solution to eq.~(\ref{eq:ModeRTE})
using this profile will not satisfy eqs.~(\ref{eq:RadEq1})-(\ref{eq:CFlux}).  To establish equilibrium, 
the initial temperature profile is corrected using the standard
Uns\"old-Lucy procedure \citep[][]{Mihalas78a}.  The entire process is iterated until 
the deviations from radiative equilibrium,  constant flux, and 
the relative size of the temperature correction are all less 
than a few percent.  During a given iteration, the specific intensity calculated 
from the previous iteration is 
used to determine the source function.  Thus, the source function must 
also converge to yield a self-consistent solution.  Numerically, we find that the source 
function converges more rapidly than the other quantities considered above.
For a more detailed discussion of the construction of 
self-consistent 
atmosphere models, see \citet[][]{HoLai01a} and \citet[][]{Mihalas78a}.

\subsection{Partial mode conversion using photon Stokes parameters}
\label{subsect:Stokes}

While the treatment described above captures the essential physics of the transfer 
problem, it is important to compare it to the exact solution obtained from integration 
of the transfer equations for the radiation Stokes parameters.
As discussed in \S\ref{sect:VacEffect}, near the vacuum resonance, the modal transfer equation (\ref{eq:ModeRTE}) 
breaks down because of the violation of the Faraday depolarization condition and collapse 
of the photon modes (see Figs. 4-5 of \citet[][]{LaiHo03a} for the precise condition).  

The radiation transfer equations for the Stokes parameters are given by \citep[][]{LaiHo03a}:
\begin{eqnarray}
\label{eq:StokesEq}
\pm\mu{\partial{\mbox{\boldmath{$I$}}}\over \partial\tau} & = &
\mbox{\boldmath{$M$}} \cdot \mbox{\boldmath{$I$}}-\mbox{\boldmath{$S_{em}$}},
\end{eqnarray}
with
\begin{eqnarray}
\mbox{\boldmath{$M$}} & = & {\omega\over c\rho\kappa_T}\left(
\begin{array}{cccc}
\sigma_{11i} & 0 & \sigma_{12i}/2 & -\sigma_{12r}/2\\
0 & \sigma_{22i} & -\sigma_{12i}/2 & -\sigma_{12r}/2\\
-\sigma_{12i} & \sigma_{12i} & (\sigma_{11i}+\sigma_{22i})/2 & (\sigma_{11r}-\sigma_{22r)/2}\\
-\sigma_{12r} & -\sigma_{12r} & (\sigma_{22r}-\sigma_{11r})/2 & (\sigma_{11i}+\sigma_{22i})/2
\end{array}
\right),\\
\mbox{\boldmath{$S_{em}$}} & = & {\omega B_{\nu}\over 2\rho\kappa_T c}\left(
\begin{array}{c}
\sigma_{11i}\\
\sigma_{22i}\\
0\\
-2\sigma_{12r}
\end{array}
\right)_{em},
\end{eqnarray}
where $\sigma_{11}=\epsilon'\cos^2\theta_{kB}+\eta'\sin^2\theta_{kB}-a$, 
$\sigma_{12}=ig\cos\theta_{kB}$, $\sigma_{22}=\epsilon'-a-m\sin^2\theta_{kB}$, 
$\sigma_{\alpha\beta r}=\Re e(\sigma_{\alpha\beta})$, $\sigma_{\alpha\beta i}=\Im m(\sigma_{\alpha\beta})$, and
$\mbox{\boldmath{$I$}}$$\equiv (I_{11},I_{22},U_{\nu},V_{\nu})^+$, with $I_{11}=(I_{\nu}+Q_{\nu})/2$, $I_{22}=(I_{\nu}-Q_{\nu})/2$.
Note that 
eq.~(\ref{eq:StokesEq}) ignores scattering; the ``em'' suffix on the source functions implies 
that terms proportional to $\gamma_{re}$ or $\gamma_{ri}$ should be set to zero, as they are related to 
scattering contributions.  The scattering contributions to eq.~(\ref{eq:StokesEq}) are derived in 
\citet[][]{LaiHo03a}.

Away from the resonance, the modes discussed in \S\ref{subsubsect:MOSF} are well defined and are 
readily calculated from the Stokes parameters.  Neglecting dissipative terms in the dielectric tensor, 
the transverse part 
of the mode polarization vectors can be written [see eq.~(\ref{eq:ModeEigenvs})]
\begin{eqnarray}
\label{eq:Epm}
\mbox{\boldmath{$e$}}_+ = (i\cos\theta_m,\sin\theta_m),\ \ \mbox{\boldmath{$e$}}_- = (-i\sin\theta_m,\cos\theta_m),
\end{eqnarray}
where $\theta_m$ is the ``mixing angle'' defined by $\cos\theta_m=K_+/\sqrt{1+K_+^2}$, $\sin\theta_m=1/\sqrt{1+K_+^2}$.  
The intensities of the $\pm$ modes can be calculated from the Stokes parameters via
\begin{eqnarray}
\label{eq:Stokes2Pm}
I_{\nu}^{\pm}={1\over 2}\left[I_{\nu}\pm\left(\cos 2\theta_m Q_{\nu} + \sin 2\theta_m V_{\nu}\right)\right].
\end{eqnarray}
Conversely, given the mode intensities, the Stokes parameters can be calculated using
\begin{eqnarray}
\label{eq:Pm2Stokes1}
I_{\nu} & = & I_{\nu}^++I_{\nu}^-,\\
Q_{\nu} & = & \cos 2\theta_m(I_{\nu}^+-I_{\nu}^-)-2\sin 2\theta_m(I_{\nu}^+I_{\nu}^-)^{1/2}\cos\Delta\phi,\\
U_{\nu} & = & -2(I_{\nu}^+I_{\nu}^-)^{1/2}\sin\Delta\phi,\\
\label{eq:Pm2Stokes4}
V_{\nu} & = & \sin 2\theta_m(I_{\nu}^+-I_{\nu}^-)+2\cos 2\theta_m(I_{\nu}^+I_{\nu}^-)^{1/2}\cos\Delta\phi,
\end{eqnarray}
where $\Delta\phi = \Delta\phi_i + (\omega/c)\int^z (n_+-n_-) dz$ is the phase difference between the 
$+$ and $-$ modes.  
Note that $\Delta\phi$ is unknown, since the initial phase difference $\Delta\phi_i$ between photons in the X 
and O-modes is random.  To 
correctly evaluate the Stokes parameters from the specific mode intensities, one should sample $\Delta\phi$ 
from a random distribution, and average over the results.  
Practically, we note that while the choice of $\Delta\phi$ affects the values of the Stokes parameters, it 
does not change the specific mode intensities calculated from eq.~(\ref{eq:Stokes2Pm}).  Therefore, the 
phase difference is unimportant for the comparison of the mode and Stokes parameter transfer equations 
(see \S\ref{subsect:Comparison}).

In principle, eq.~(\ref{eq:StokesEq}) can be integrated from $\tau_{max}$ to $\tau_{min}$ using the 
initial condition $\mbox{\boldmath{$I$}}(\tau_{max})=(B_{\nu}/2,B_{\nu}/2,0,0)^+$ as in \S\ref{subsect:Modes}.
However, this approach runs into a numerical difficulty:
away from the vacuum resonance, differences in 
the indices of refraction for 
the two modes manifest as rapid oscillations in $Q_{\nu}$, $U_{\nu}$, $V_{\nu}$, which are difficult to 
handle numerically.
Thus, the direct 
solution of eq.~(\ref{eq:StokesEq}) over the entire range of integration is impractical.  
It is possible, however, to integrate eq.~(\ref{eq:StokesEq}) for a small range of $\tau$ around the resonance.
Using eqs. (\ref{eq:Stokes2Pm}) and (\ref{eq:Pm2Stokes1})--(\ref{eq:Pm2Stokes4}), we can quantitatively 
compare the result of such an integration with that obtained using the method of 
\S\ref{subsubsect:ModeSol}, and thereby 
confirm the 
accuracy of the latter method (see \S\ref{subsect:Comparison}).


\subsection{Numerical comparison between mode and Stokes transfer equations}
\label{subsect:Comparison}

\begin{figure}
\includegraphics[width=84mm]{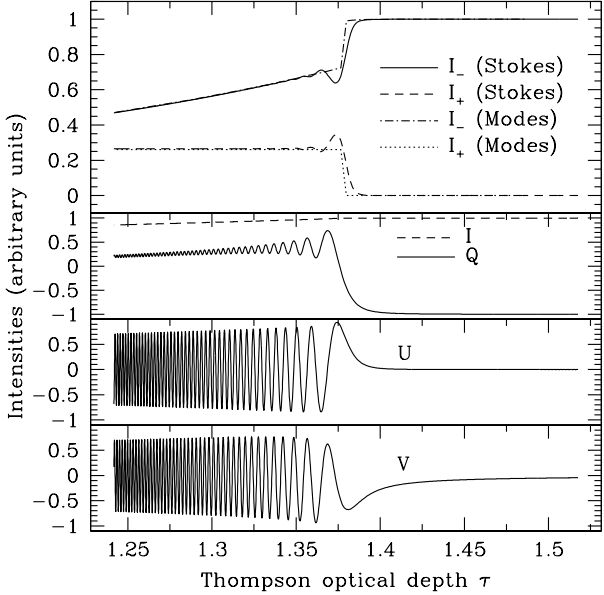}
\caption{Evolution of the Stokes parameters across the vacuum resonance obtained by integrating the transfer 
equation~(\ref{eq:StokesEq}).  The parameters are $B=10^{14}$ G, $\theta_{kB}=\pi/4$, $T=5\times 10^6$ K, 
and $E=1$ keV.  At high optical depth, the photon is in the $-$ mode.  In the top panel, the dotted line and 
the dot-dashed line show the mode intensities obtained using the method described in \S\ref{subsect:Modes} 
[i.e., solving the transfer equation (\ref{eq:ModeRTE}) based on photon modes, but taking account of partial 
mode conversion through eqs.~(\ref{eq:ModeConv2})-(\ref{eq:ModeConv3})],
while the solid and dashed lines give the mode intensities based on the evolution of Stokes parameters 
(panels 2-4).
Note the close agreement everywhere except near the resonance where the modes are not well-defined.}
\label{fig:Comparison}
\end{figure}

We consider a typical case, the propagation of a photon, 
initially polarized in the $-$ mode, with energy 
$E=1.0$ keV, propagation angle $\theta_{kB}=\pi/4$, 
and magnetic field $B=10^{14}$ G.  The temperature profile is 
held constant at $T=5\times 10^{6}$ K
[eqs.~(\ref{eq:Pm2Stokes1})-(\ref{eq:Pm2Stokes4}) 
are used to set the 
initial conditions for (\ref{eq:StokesEq})].  
Figure~\ref{fig:Comparison} shows the Stokes parameters as a function of optical depth near the resonance.
These are obtained by integrating eq.~(\ref{eq:StokesEq}).
The corresponding mode intensities 
are then calculated using eq.~(\ref{eq:Stokes2Pm}) and depicted in the top panel (solid and dashed lines).
The dashed-dot and dotted lines show the results obtained from the integration of the mode equations 
with partial mode conversion [eqs.~(\ref{eq:ModeConv2})-(\ref{eq:ModeConv3})]. 
Note that the curves agree exactly except near the resonance where the modes are not well-defined.

We have carried out many similar comparisons between the mode equations and the Stokes transfer equation.  The 
close agreement between the two methods establishes the validity of our method described in \S\ref{subsect:Modes},
i.e., integrating the mode eq.~(\ref{eq:ModeRTE}) and taking partial mode conversion into account using 
eqs.~(\ref{eq:ModeConv2})-(\ref{eq:ModeConv3}).


\subsection{Numerical grid and test of accuracy}
\label{subsect:GridTests}

In solving eq.~(\ref{eq:ModeRTE}), we set up grids in 
Thomson optical depth, temperature, density, energy, and angle.  
The grid in optical depth
$\{\tau_d:d=1,\ldots,D\}$ is equally spaced logarithmically 
with $15-20$ points per decade (ppd).
As discussed above, this grid spans $5-8$ orders of magnitude 
to insure that photons are generated at densities higher 
than the X-mode decoupling depth, 
and that they are fully decoupled from the matter at the outermost layer.

Care must be used in defining the energy grid.
As mentioned before, 
vacuum polarization  
introduces a narrow spike in the X 
mode opacity.  A prohibitively high energy 
grid resolution is required to properly resolve this 
feature.  An alternative is to use the equal-grid method 
described by \citet[][]{HoLai03a}.  In this case, each 
point of the energy grid is chosen to be the vacuum 
resonance energy [given by eq.~(\ref{eq:E_V})] 
corresponding to a point on the optical depth grid:
$\{E_n=E_V(\tau_n):n=1,\ldots,D\}$.  This 
insures that the vacuum resonance is resolved.
\citet[][]{HoLai03a} point out that in the ``no conversion'' limit, this leads to an 
over-estimate of the integrated optical depth 
across the vacuum resonance.  It is therefore 
important to investigate what effect this has on the 
emergent spectra.

\begin{figure}
\includegraphics[width=84mm]{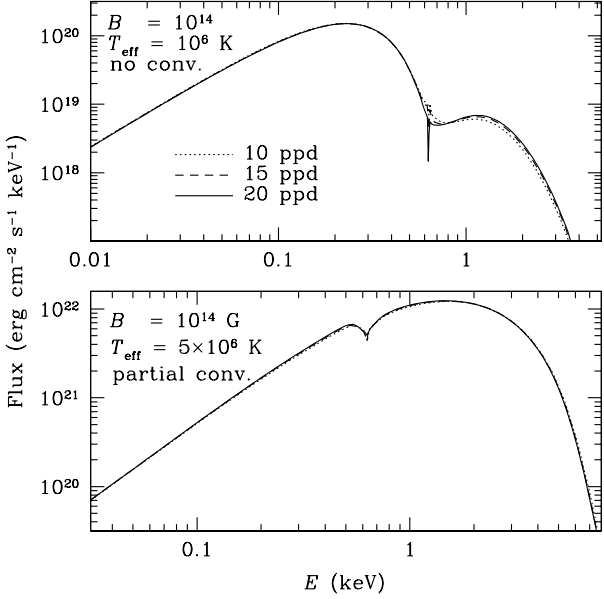}
\caption{Spectra showing the effect of grid resolution on thermal emission.  The top panel shows hydrogen 
atmosphere models 
with $B=10^{14}$ G, $T_{\rm eff}=10^6$ K, which include vacuum polarization but neglect mode conversion.  As the 
number of grid points per decade (ppd) are increased, the curves quickly converge to the 20 ppd case.  At low resolution 
the error mainly occurs around the proton cyclotron feature, and is negligible elsewhere.  The bottom 
panel shows hydrogen models with $B=10^{14}$ G, $T_{\rm eff}=5\times 10^6$ K, which include vacuum polarization and 
mode conversion.  At higher effective temperature, the difference between models with varying grid resolution 
is negligible.}
\label{fig:GridConv}
\end{figure}

Figure~\ref{fig:GridConv} illustrates the effect of grid 
resolution on the spectra for two of the models presented in 
\S\ref{sect:Results}.  The top panel shows the model with $B=10^{14}$ G,
$T_{\rm eff}=10^6$ K, which 
includes vacuum polarization in the opacities, but neglects the mode conversion 
effect (i.e., $P_{\rm jump}=1$).  Over-estimation of the vacuum resonance in the 
X-mode opacity is expected to be strongest for this 
model, since modification of the emission spectrum is 
due solely to the enhaced opacity.  The difference 
between models at $15$ ppd and $20$ ppd is negligible.  Even at 
$10$ ppd the difference is small, occuring mainly around the proton 
cyclotron line, as expected.

The bottom panel of Fig.~\ref{fig:GridConv} shows the model with $B=10^{14}$ G, and 
$T_{\rm eff}=5\times 10^6$ K, which includes partial mode conversion.  
At higher effective temperatures, the optical depth across the vacuum 
resonance becomes much greater than unity, thus, the error due to finite 
grid size becomes even less important.  The lower grid resolution ($10$ ppd) model
shows negligible deviation from the higher-grid resolution (20 ppd) model, even around the proton cyclotron 
feature.  This behavior is typical of all models with 
$B=5\times 10^{14}$ G.


\section{Results: Atmosphere Structure, Spectra, and Emission Beam Pattern}
\label{sect:Results}

We now present the results of our atmosphere models.  We consider $B=4\times 10^{13}, 7\times 10^{13}, 
10^{14}, 5\times 10^{14}$ G, and $T_{\rm eff}=10^6, 5\times 10^6$ K, for both H and He compositions.


\subsection{Atmosphere structure}
\label{subsect:AtmoStruct}

\begin{figure}
\includegraphics[width=84mm]{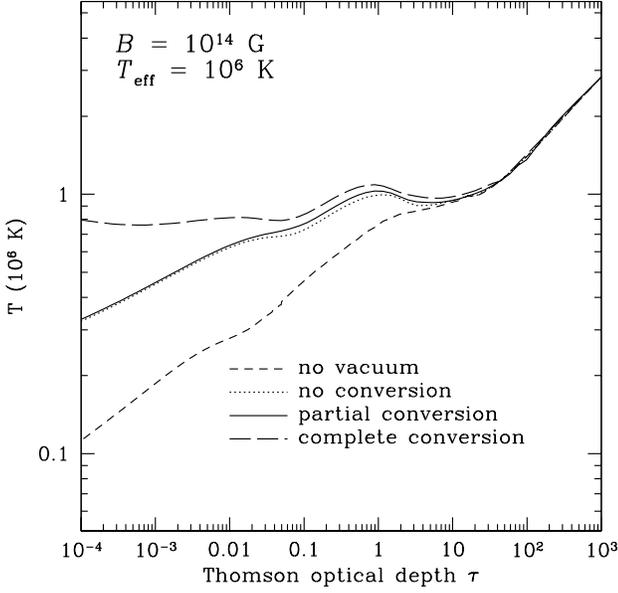}
\caption{Temperature profile for the hydrogen atmosphere model with $B=10^{14}$ G, $T_{\rm eff}=10^6$ K.  The four 
curves correspond to different ways of treating the vacuum polarization effect:
(1) no vacuum (short-dashed curve); (2) no conversion (dotted curve); 
(3) partial conversion (solid curve); and (4) complete conversion (long-dashed curve).}
\label{fig:Temp10}
\end{figure}

Figure~\ref{fig:Temp10} shows the temperature profile 
as a function of Thompson optical depth $\tau$ for the H atmosphere model with 
$B=10^{14}$ G, and $T_{\rm eff}=10^6$ K.  
To understand the effect of vacuum polarization, we show the results based on 
four different ways of treating vacuum polarization: (1) vacuum polarization effect 
is completely turned off (``no vaccum''); (2) vacuum polarization is included, but 
the mode conversion is neglected ($P_{\rm jump}=1$, ``no conversion''); (3) vacuum polarization 
is included, and complete mode conversion is assumed ($P_{\rm jump}=0$, ``complete conversion''); 
(4) vacuum polarization is included with the correct treatment of the resonance 
(``partial conversion''), using $P_{\rm jump}$ 
calculated from eq.~(\ref{eq:P_j}).
We see that models which include vacuum polarization 
show higher temperatures over a wide range of $\tau$ 
for the same $T_{\rm eff}$ than models which ignore vacuum effects.
This temperature increase 
is due to the X-mode opacity feature at the vacuum resonance (see Fig.~\ref{fig:Kappa}).
In general, 
atmosphere structure is determined by the radiative equilibrium condition, 
and 
inspection of  the individual terms of eq.~(\ref{eq:RadEq1}) reveals 
how the resonance affects the temperature profile.  The mode 
absorption opacities obey the relation $\kappa_O\gg\kappa_X$ 
except at the energies $E=E_{Bi},E_V$.  Thus, $\kappa_X(B_{\nu}/2)$ 
can be neglected relative to $\kappa_O(B_{\nu}/2)$ in eq.~(\ref{eq:RadEq1}).  
In the absence of vacuum polarization, the O mode largely 
determines the atmosphere structure due to the weak interaction 
of X mode photons with the medium.  However, when the resonance 
spike in the X mode opacity is present, $\kappa_X J^X_{\nu}$ cannot 
be neglected relative to $\kappa_O J^O_{\nu}$; in fact, this occurs 
over a large bandwidth for which $J^X_{\nu}\gg J^O_{\nu}$.  The 
result of this enhanced interaction is to increase the overall temperature.
Adding the effect of mode conversion further increases the temperature 
over a large range of optical depth.  This is due to heat deposited 
by converted X mode photons, which interact with the large O mode 
opacity after passing through the vacuum resonance. 
The temperature profile for the partial mode conversion 
model (shown by the solid curve of Fig.~\ref{fig:Temp10}) closely follows the result for 
the no conversion model (shown by the dotted curve) for the small optical depths at which low 
energy photons decouple.  This is because for these photons, $E\la E_{ad}$, is satisfied and 
mode conversion is ineffective.  
For larger optical depths, 
at which higher energy photons decouple, $E\ga E_{ad}$, and mode conversion is 
more effective, thus the ``partial conversion'' result lies between the ``no conversion'' and 
``complete conversion'' limits.

\begin{figure}
\includegraphics[width=84mm]{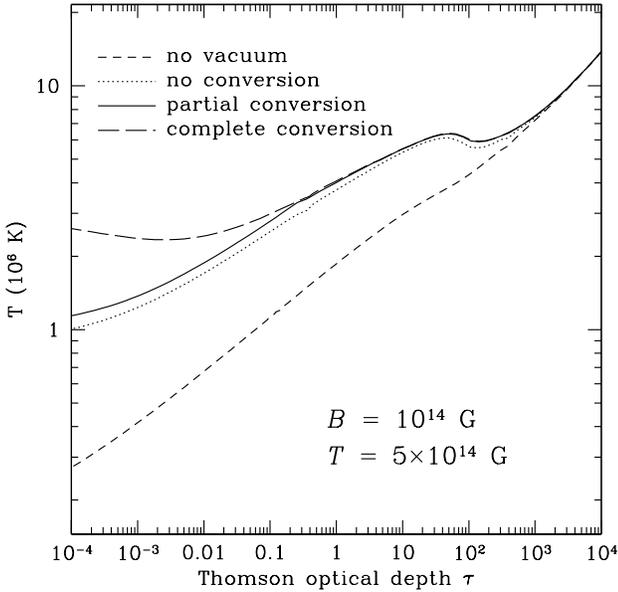}
\caption{Same as Fig.~\ref{fig:Temp10}, except for $T_{\rm eff}=5\times 10^6$ K.}
\label{fig:Temp50}
\end{figure}

Figure~\ref{fig:Temp50} shows the temperature profile for the $B=10^{14}$ G, $T_{\rm eff}=5\times 10^6$ K 
model.
This higher temperature model shows the same basic features as 
the low-$T_{\rm eff}$ model in 
Fig.~\ref{fig:Temp10}.  In this case,
the energy flux is carried by photons with higher energies, and the adiabatic condition ($E\ga E_{ad}$) is more 
readily satisfied, leading to effective mixing of photon modes.  Thus we see that at large optical 
depths ($\tau\ga 0.1$), the 
partial conversion profile closely follows the complete conversion
curve.  At lower optical depth, the partial conversion 
profile lies between the complete conversion and 
no conversion curves.

Finding self-consistent temperature profiles is the most time consuming step in atmosphere modeling.  
Once the profile is 
known, the emergent radiation can be obtained by a single integration of the transfer equation.  
To facilitate future work on NS 
atmospheres and related applications, we provide fitting formulae for the models presented in this paper.
Formulae are provided only for models incorporating vacuum polarization with partial mode conversion.
The fits are valid over the optical depth range $\tau=10^{-3}-2\times 10^{4}$.  Each model is fit by the function
\begin{eqnarray}
\log_{10}\left [T_6(\tau)\right ] & = & \left\{
\begin{array}{cc}
a_1+a_2\,\Delta x+a_3\,\Delta x^2+a_4\,\Delta x^3+a_5\,\Delta x^4 + a_6\,\Delta x^5 & 
\tau_{\rm mid} < \tau < 2\times 10^4,\\
a_1+a_2\,\Delta x+b_3\,\Delta x^2+b_4\,\Delta x^3+b_5\,\Delta x^4 + b_6\,\Delta x^5 & 
10^{-3} < \tau < \tau_{\rm mid},
\end{array}\right.
\label{eq:tprofile}
\end{eqnarray}
where $x\equiv\log_{10}(\tau)$, $\Delta x\equiv x-x_{\rm mid}$, 
and $\tau_{\rm mid}$ denotes the break between the two parts of the fit necessary to 
describe the temperature profile.  The parameters for each model are summarized in Table~\ref{Table:Fitting}.

\begin{table}
\caption{The parameters in the fitting formulae [Eq.~(\ref{eq:tprofile})]
for the temperature profiles for atmosphere models
with different magnetic field strengths, effective temperatures and 
compositions (ionized H or He)}
\label{Table:Fitting}
\begin{tabular}{lccccccc}
\hline
Model & $\tau_{\rm mid}$ & $a_1$ & $a_2$ & $a_3$ & $a_4$ & $a_5$ & $a_6$ \\
&&&& $b_3$ & $b_4$ & $b_5$ & $b_6$ \\
\hline
$10^{13}$ G, $5\times 10^6$ K, H & 27.1 & 0.793 & 0.122 & -0.502 & 0.548 & -0.205 
	& 0.0266 \\
&&&& 0.00445 & 0.0108 & 0.00211 & 0.0000574 \\
$4\times 10^{13}$ G, $10^6$ K, H & 4.27 & -0.0599 & 0.192 & 0.0225 & 0.0115 & -0.0072 
	& 0.00116 \\ 
&&&& 0.109 & 0.0828 & 0.0256 & 0.00286 \\
$4\times 10^{13}$ G, $5\times 10^6$ K, H & 11.9 & 0.623 & -0.0425 & 0.0991 & 0.0412 & -0.026 
	& 0.0036 \\ 
&&&& -0.0851 & -0.00392 & 0.0034 & 0.000418 \\
$7\times 10^{13}$ G, $10^6$ K, H & 0.888 & -0.0455 & -0.158 & 0.221 & -0.0469 & 0.00231 
	& 0.000307 \\ 
&&&& -0.329 & -0.118 & -0.00387 & 0.00304 \\
$7\times 10^{13}$ G, $5\times 10^6$ K, H & 21.6 & 0.789 & 0.123 & -0.650 & 0.726 & -0.274 
	& 0.0354 \\
&&&& -0.0105 & -0.00406 & -0.00242 & -0.000411 \\
$10^{14}$ G, $10^6$ K, H & 0.683 & 0.00828 & 0.0614 & -0.304 & 0.266 & -0.0719 
	& 0.00652 \\
&&&& -0.313 & -0.374 & -0.162 & -0.0234 \\
$10^{14}$ G, $10^6$ K, He & 0.749 & -0.0935 & -0.154 & 0.262 & -0.0904 & 0.0167 
	& -0.00124 \\
&&&& -0.197 & 0.0197 & 0.0428 & 0.0083 \\
$10^{14}$ G, $5\times 10^6$ K, H & 30.6 & 0.799 & 0.115 & -0.537 & 0.617 & -0.241 
	& 0.0326 \\
&&&& -0.00603 & 0.00409 & 0.000351 & -0.0000978 \\
$5\times 10^{14}$ G, $10^6$ K, H & 32.9 & 0.0939 & 0.0181 & -0.0153 & -0.0413 & 0.0376 & -0.00578 
	\\
&&&& 0.0504 
& 0.0462 & 0.00991 & 0.000676 \\
$5\times 10^{14}$ G, $5\times 10^6$ K, H & 63.2 & 0.761 & 0.00198 & 0.267 & -0.356 & 0.179 
	& -0.0282 \\
&&&& -0.118 & -0.0336 & -0.00428 & -0.00022 \\
$5\times 10^{14}$ G, $5\times 10^6$ K, He & 23.5 & 0.707 & 0.0467 & 0.342 & -0.417 & 0.174 
	& -0.0234 \\
&&&& -0.109 & -0.040 & -0.0069 & -0.000481 \\
\hline
\end{tabular}

\end{table}


\subsection{Spectra}
\label{subsect:Spectra}

\begin{figure}
\includegraphics[width=84mm]{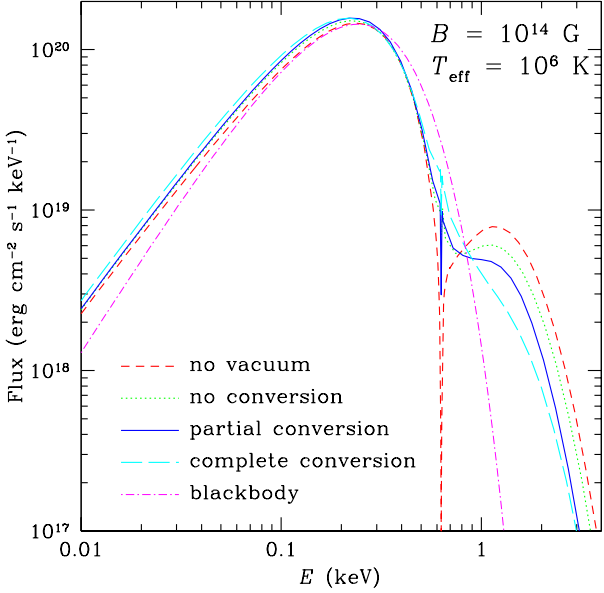}
\caption{Spectra for hydrogen atmosphere models with $B=10^{14}$ G, $T_{\rm eff}=10^6$ K.  The four cases described 
in the text are shown: (1) no vacuum (short-dashed cuve); (2) no conversion (dotted curve); (3) partial conversion 
(solid curve); and (4) complete conversion (long-dashed curve).  The light dashed-dot curve shows the blackbody 
spectrum with $T=10^6$ K.
For all three cases that include vacuum 
effects, the proton cyclotron feature is strongly suppressed and the high energy tail is softened relative to the 
no vacuum case.  The ``partial conversion'' curve is seen to be intermediate between the ``no conversion'' and 
``complete 
conversion'' limits.}
\label{fig:Specb10t10}
\end{figure}

Figure~\ref{fig:Specb10t10} presents the spectrum for the hydrogen 
atmosphere model with $B=10^{14}$~G and $T_{\rm eff}=10^6$~K.  The results for the four different 
ways of treating 
vacuum polarization (see \S\ref{subsect:AtmoStruct}) are shown.  These spectra correspond to the temperature profiles 
depicted in Fig.~\ref{fig:Temp10}.

When the vacuum polarization effect is neglected, the spectrum of a magnetic, ionized H atmosphere model generally 
exhibits two characteristics: (1) a hard spectral tail (compared to blackbody) due to the non-grey free-free opacity 
\citep[$\kappa^{\rm ff}$ decreases with increasing photon energy;][]{Shibanovetal92a,Pavlovetal95a};
(2) a significant 
proton cyclotron absorption line when $E_{Bi}$ is not too far away from the blackbody 
peak ($\sim 3 k_B T$) 
\citep[][]{Zaneetal01a,HoLai01a}.  Vacuum polarization tends to soften the hard spectral tail and suppress 
(reduce) the proton 
cyclotron line.
These effects 
are discussed extensively in \S 4 of \citet[][]{HoLai03a}, and \citet[][]{LaiHo03a}.
In Fig.~\ref{fig:Specb10t10}, all of the models that include vacuum polarization effects 
display a large reduction in the 
equivalent width (EW) of the proton cyclotron feature due to the modification of the temperature profile 
by the vacuum resonance (see \S\ref{subsect:AtmoStruct}) and the mode conversion effect.
The spectra also show softening of the hard spectral tail relative to 
the no vacuum case, though they are all still harder than 
blackbody.  The ``partial 
conversion'' curve appears as an intermediate case between 
the ``complete conversion'' and ``no conversion'' extremes.  The 
adiabatic regime where mode conversion is efficient is clearly 
visible: for $E\ga E_{ad}\sim 2$ keV, the ``partial conversion'' 
curve begins to follow the ``complete conversion'' curve. 

\begin{figure}
\includegraphics[width=84mm]{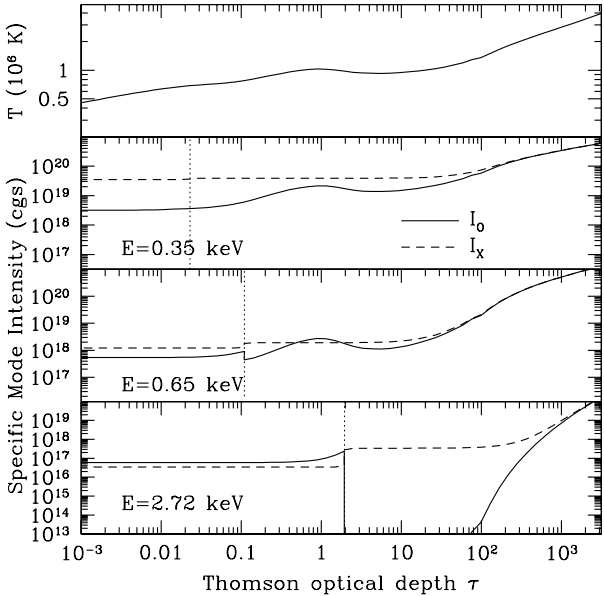}
\caption{The transition from ``no conversion'' to ``complete conversion'' for a photon propagating in a hydrogen 
atmosphere 
with $B=10^{14}$ G, $T_{\rm eff}=10^6$ K. 
The top panel shows the temperature profile for this model.
The bottom three panels show the evolution of the specific mode intensities 
for energies $E=0.35, 0.65, 2.72$ keV and $\theta_{kB}=\pi/4$.  
The dashed line shows the X-mode intensity, the solid line the O-mode intensity, and 
the dotted vertical lines specify the 
location of the vacuum resonance.  For $E=0.35$ keV: $H_{\rho}=0.69$ cm, $E<E_{ad}$, and $P_{\rm jump}=1.0$, 
leading to minimal mode conversion.  For $E=0.65$ keV: $H_{\rho}=0.78$ cm, $E\sim E_{ad}$, and 
$P_{\rm jump}=0.65$, leading to partial mixing of the modes.  For $E=2.72$ keV: $H_{\rho}=0.98$ cm, $E>E_{ad}$, and 
$P_{\rm jump}=0.15$, leading to nearly complete conversion of the modes.}
\label{fig:Tauint}
\end{figure}

This transition from ``no conversion'' to ``complete conversion'' is further illustrated 
by Fig.~\ref{fig:Tauint}, which shows, for several photon energies $E$ 
(and a given
direction of propagation, $\theta_{kB}$)
the evolution of the specific mode intensities as a function of optical depth.  
At the resonance depth (denoted by the vertical lines), the X mode photons 
encounter the vacuum induced spike in opacity, and 
mode conversion occurs [governed by the non-adiabatic jump probability 
$P_{\rm jump}$; see eq.~(\ref{eq:P_j})].  As the energy is increased (from the second to bottom 
panels), mode conversion becomes increasingly more effective.

\begin{figure}
\includegraphics[width=84mm]{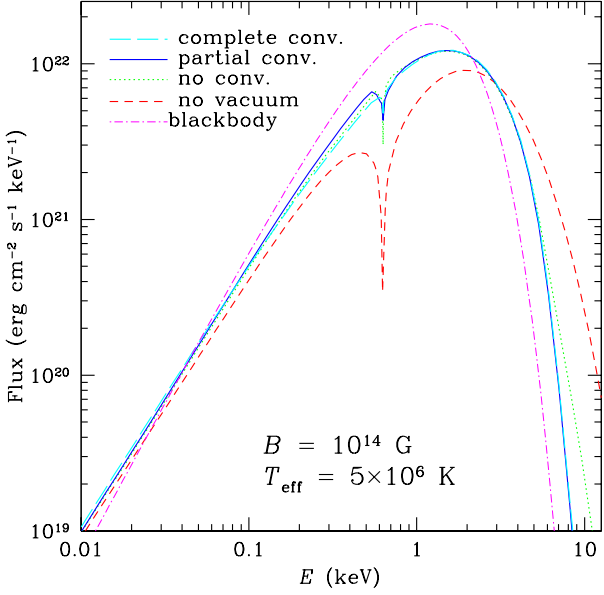}
\caption{Same as Fig.~\ref{fig:Specb10t10}, except $T_{\rm eff}=5\times 10^6$ K.  The strength of the proton 
cyclotron feature is strongly suppressed for models which include vacuum polarization.  At $E\ga 5$ keV, the 
models which include mode conversion are softer than those without, though all the atmosphere models are still 
harder than blackbody.}
\label{fig:Specb10t50}
\end{figure}

Figure~\ref{fig:Specb10t50} shows the spectrum for the $B=10^{14}$ G and 
$T_{\rm eff}=5\times 10^6$ K hydrogen model.  All the calculations that include
vacuum effects show strong suppression of the 
ion cyclotron feature and significant softening of the 
hard spectral tail.
At higher effective temperatures, there is a smaller 
difference between the no conversion, partial conversion, 
and complete conversion cases.  In this regime, the optical depth 
across the vacuum resonance is much greater than unity, and the 
decoupling of X-mode photons occurs at the resonance density whether or not 
mode conversion is taken into account \citep[see][]{LaiHo02a}.
At high energies ($E\ga 5$ keV), the 
models which include mode conversion are softer 
than those which do not.

\begin{figure}
\includegraphics[width=84mm]{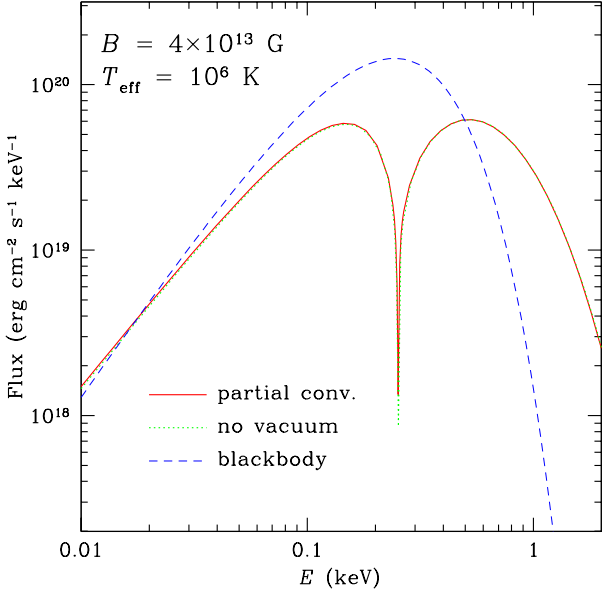}
\caption{Spectrum of the hydrogen atmosphere model with $B=4\times 10^{13}$ G, $T_{\rm eff}=10^6$ K, calculated for 
two cases: partial mode conversion (solid curve), and no vacuum (dotted curve).  The light dashed line shows 
the blackbody spectrum with $T=10^6$ K.}
\label{fig:Specb4t13}
\end{figure}

\begin{figure}
\includegraphics[width=84mm]{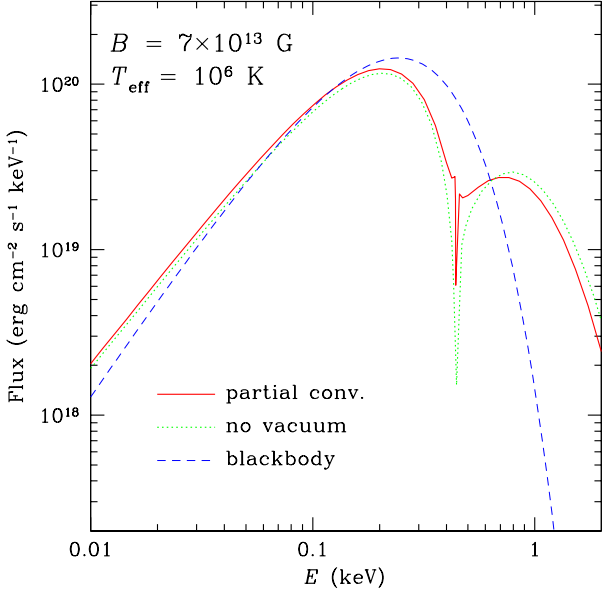}
\caption{Same as Fig.~\ref{fig:Specb10t10}, except for $B=7\times 10^{13}$ G.} 
\label{fig:Specb7t13}
\end{figure}

\begin{figure}
\includegraphics[width=84mm]{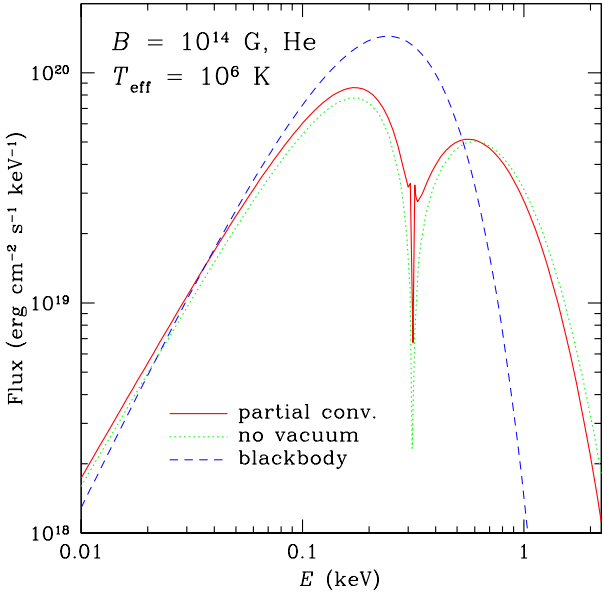}
\caption{Same as Fig.~\ref{fig:Specb4t13}, except for helium composition with $B=10^{14}$ G.  
The ion cyclotron line width is reduced by vacuum polarization, though it has a larger equivalent width than the 
model for hydrogen, due to the location of the line near the maximum of the continuum emission.}
\label{fig:Specheb1t10}
\end{figure}

\begin{figure}
\includegraphics[width=84mm]{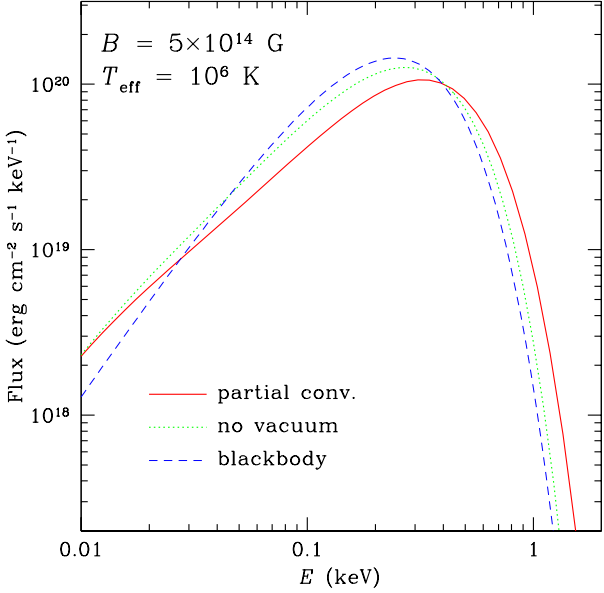}
\caption{Same as Fig.~\ref{fig:Specb7t13}, except with $B=5\times 10^{14}$ G.}
\label{fig:Specb50t10}
\end{figure}

Figures~\ref{fig:Specb4t13}-\ref{fig:Specb50t10} depict 
atmosphere models at $T_{\rm eff}=10^6$ K with varying 
magnetic field strengths, 
comparing the ``no vacuum'' and correct ``partial conversion'' results.
For $B=4\times 10^{13}$ G $<B_l$ (Fig.~\ref{fig:Specb4t13}),  
the vacuum resonance lies at a lower density than the X-mode and O-mode photospheres, and 
vacuum polarization has a negligible effect on the spectrum, reflected by the close agreement between the 
``no vacuum'' and ``partial conversion'' curves.
For the hydrogen atmosphere model with $B=7\times 10^{13}$ G (Fig.~\ref{fig:Specb7t13}), and 
the helium atmosphere model with $B=10^{14}$ G (Fig.~\ref{fig:Specheb1t10}), 
vacuum polarization affects the spectrum.  For these models, $B\ga B_l$, and the ion cyclotron line lies 
near the blackbody peak. 
Thus, it is particularly important to treat the vacuum resonance correctly, taking partial mode conversion into 
account to 
calculate the line width.  For the $B=5\times 10^{14}$ G model (Fig.~\ref{fig:Specb50t10}), 
the spectral feature at $E_{Bi}$ is outside the energy band of observational interest.  We note that at such a high 
field and low effective temperature, the atmosphere should contain a significant fraction of bound atoms and 
molecules \citep[][]{Hoetal03a,Potekhinetal04a}, so the fully ionized model shown in 
Fig.~\ref{fig:Specb50t10} is not realistic.

\begin{figure}
\includegraphics[width=84mm]{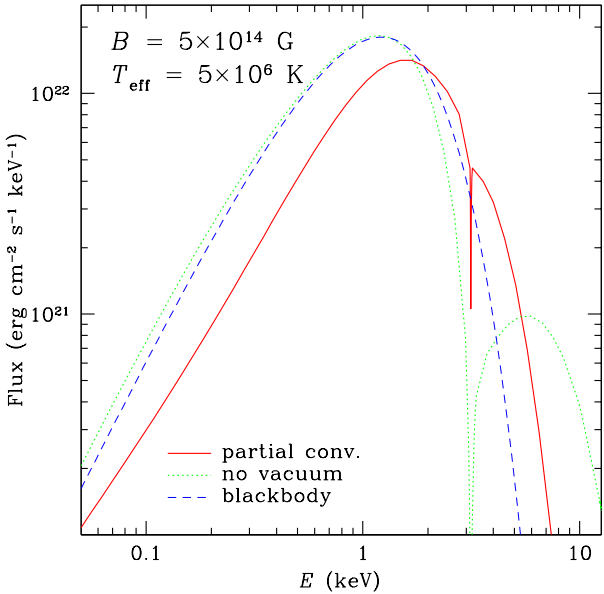}
\caption{Same as Fig.~\ref{fig:Specb50t10}, except for $T_{\rm eff}=5\times 10^6$ K.  The proton cyclotron is strongly 
suppressed by vacuum polarization.  The hard spectral tail is softened considerably relative to the no vacuum 
case, though it is still harder than blackbody.}
\label{fig:Specb5t14}
\end{figure}

\begin{figure}
\includegraphics[width=84mm]{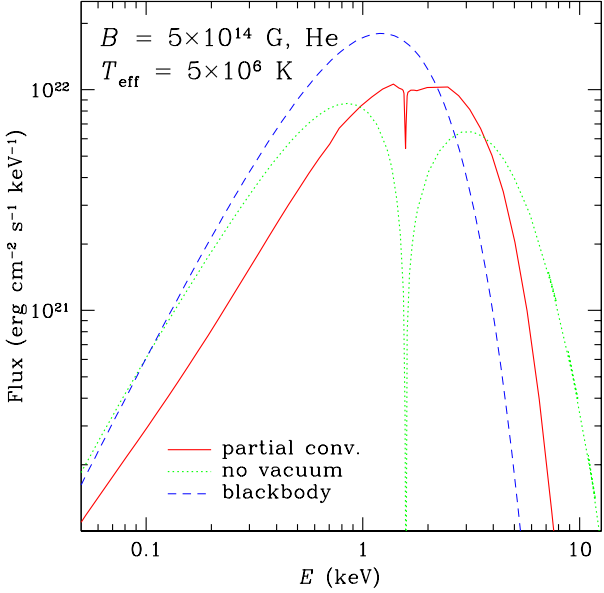}
\caption{Same as Fig.~\ref{fig:Specb5t14}, except for helium composition.  Vacuum polarization strongly suppresses 
the ion cyclotron feature, and softens the hard spectral tail relative to the no vacuum case.}
\label{fig:Specb5t14he}
\end{figure}

Figures~\ref{fig:Specb5t14}-\ref{fig:Specb5t14he} show 
atmosphere models at magnetic field strength 
$B=5\times 10^{14}$ G and $T_{\rm eff}=5\times 10^6$ K, for 
hydrogen and helium compositions.  At this effective 
temperature, the ion cyclotron feature lies close to the blackbody peak, and 
the effects of vacuum polarization on the line width and the spectral tail are rather 
pronounced.

\subsection{Emission beam pattern}
\label{subsect:BeamPattern}

\begin{figure}
\includegraphics[width=84mm]{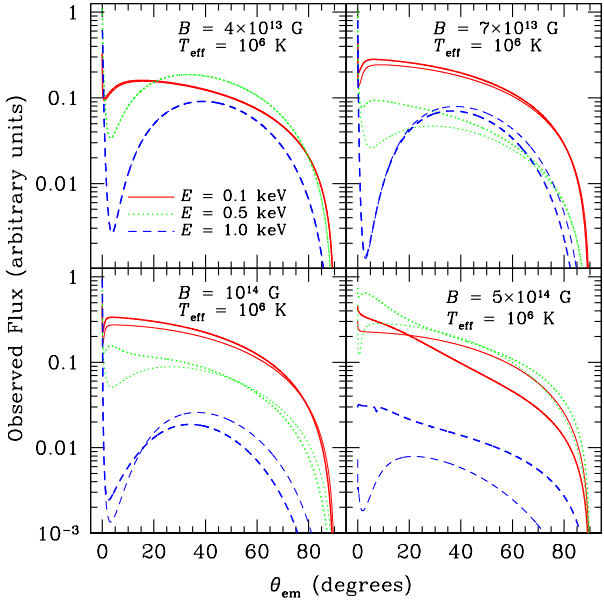}
\caption{The specific intensity from a local patch of NS as function of emission angle 
$\theta_{\rm em}$ ($=\cos^{-1}\mu, $ the angle between ${\bf\hat{k}}$ and the surface normal)
for hydrogen models with 
$T_{\rm eff}=10^6$ K, and 
$B=4\times 10^{13}$ G, $7\times 10^{13}$ G, $10^{14}$ G, $5\times 10^{14}$ G.  The heavy curves 
show models that include vacuum polarization effects, while the lightcurves show models that neglect vacuum 
polarization.  The latter models exhibit a characteristic beaming pattern, with a thin ``pencil'' shape at low 
emission angles and a broad ``fan'' at large emission angles.  Inclusion of vacuum polarization tends to 
reduce the gap and lead to a broad, featureless beaming pattern at large magnetic fields.}
\label{fig:beaming}
\end{figure}

Calculations of observed NS lightcurves and 
polarization signals (see \S\ref{sect:LCurvesPolar}) 
are critical for interpreting observations.  An important ingredient of such calculations 
is the angular beam pattern of surface emission.  Figure~\ref{fig:beaming} shows
the radiation intensity 
from a local patch of NS atmosphere (for the $T_{\rm eff}=10^6$ K models presented in \S\ref{subsect:Spectra})
as a function of emission angle relative to the 
surface normal at  
several photon energies.  The heavy curves 
show models that include vacuum polarization effects, while the lightcurves show models 
that neglect vacuum polarization.  Magnetized atmosphere models which neglect vacuum polarization have a 
distinctive beaming pattern, consisting of a thin ``pencil'' feature at low emission angles and a broad 
``fan'' beam at large emission angles, with a prominent gap between them
\citep[e.g., c.f.,][]{Ozel01a}.  This gap tends to increase with increasing photon 
energy.  The detailed shape of the emission beam pattern is determined by the temperature profile 
and the anisotropy of the mode opacities.
As shown by Fig.~\ref{fig:beaming}, vacuum polarization tends to smooth out the gap, leading 
to a broad, featureless beam pattern at large magnetic fields.
The broadening of the 
beaming pattern is due to the alteration of the temperature 
profile by the spike in opacity at the vacuum resonance and the mode conversion effect.

\begin{figure}
\includegraphics[width=84mm]{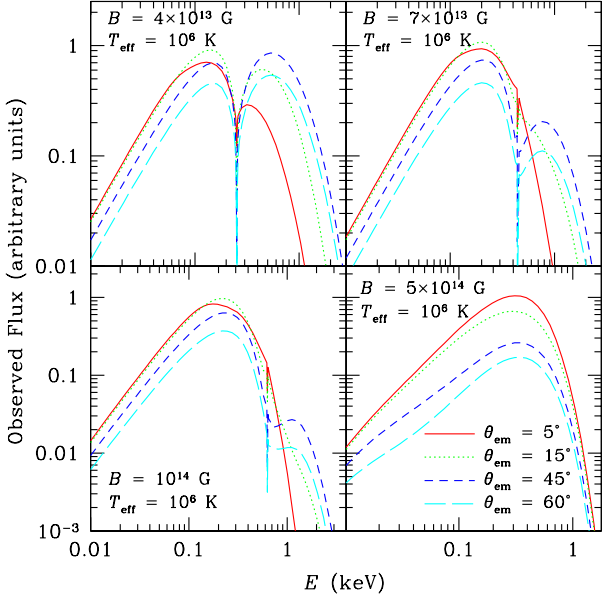}
\caption{The specific radiation intensity 
from a local patch of NS atmosphere at several emission angles for hydrogen models 
with $T_{\rm eff}=10^6$ K, and 
$B=4\times 10^{13}$ G, $7\times 10^{13}$ G, $10^{14}$ G, and $5\times 10^{14}$ G.  The shape of the 
spectrum and EW of the absorption feature change significantly with emission angle.}
\label{fig:spectra}
\end{figure}

Figure~\ref{fig:spectra} 
shows the specific radiation intensity from a patch of NS atmosphere (for the $T_{\rm eff}=10^6$ K models presented in \S\ref{subsect:Spectra})
at several emission angles.  We see that the shape of the spectrum and EW of the absorption feature can change 
significantly depending on 
the emission angle and whether or not vacuum polarization effects are included in the calculation.

\section{Polarization of the Atmosphere Emission}
\label{sect:LCurvesPolar}

Thermal emission from a magnetized NS atmosphere is highly polarized.
This arises from that fact that the typical X-mode photon opacity
is much smaller than the O-mode opacity,\footnote{This applies
under typical conditions, when the photon energy $E$ is much less than 
the electron cyclotron energy $E_{Be}$, 
is not too close to the ion cyclotron energy 
$E_{Bi}$, the plasma density is not too close to 
the vacuum resonance (see the text), and $\theta_{kB}$ (the angle between 
${\bf k}$ and ${\bf B}$) is not close to $0^\circ$ or $180^\circ$.}
$\kappa_{\rm X}\sim (E/E_{Be})^2\kappa_{\rm O} \ll \kappa_{\rm O}$.
Thus, X-mode photons escape from deeper, hotter layers of the NS atmosphere 
than the O-mode photons, and the emergent radiation is linearly 
polarized to a high degree (as high as $100\%$) 
\citep[see, e.g.,][]{GnedinSunyaev74a,Meszaros88a,PavlovZavlin00a}

There has been some recent interest in X-ray polarimetry for NSs
\citep[][]{Costaetal01a,Costaetal06a}.  
Observations of X-ray polarization, particularly when phase-resolved
and measured in different energy bands, can provide useful constraints
on the magnetic field strength and geometry, the NS rotation
rate and compactness. This will be highly complimentary 
to the information obtained from the spectra and lightcurves.
Moreover, as we show below (see also Lai \& Ho 2003b), 
vacuum resonance gives rise to a unique polarization signature in the 
surface emission, even for NSs with ``normal'' ($B\sim 10^{12}-10^{13}$~G)
magnetic fields. 

Below, we calculate the observed X-ray polarizaztion signals in 
the case when the emission comes from a rotating magnetic 
hotspot on the NS surface. Although this represents the simplest situation,
it captures the essential properties of the polarization signals, and the result
can be carried over to more general situations. See
the end of \S \ref{subsect:CircPol}
for a discussion of the case when emission
comes from an extended area (or the whole) of the stellar surface.

\subsection{Geometry and lightcurves}
\label{subsect:LCurves}

\def\bmu{{\mbox{\boldmath $\mu$}}}

To calculate the observed lightcurves and polarization signals, we
set up a fixed coordinate system $XYZ$ with the $Z$-axis along the
line-of-sight (pointing from the NS toward the observer), 
and the $X$-axis in the plane spanned
by the $Z$-axis and ${\bf\Omega}$ (the spin angular velocity
vector). The angle between ${\bf\Omega}$ and ${\bf\hat{e}}_Z$ is denoted by
$\gamma$.  The hotspot is assumed to be at the intersection of the NS
surface and dipole magnetic axis, which is inclined at an angle $\eta$
relative to the spin axis.  As the star rotates, the angle $\Theta$ 
between the magnetic axis $\bmu$ and the line of sight varies according to
\be
\cos\Theta=\cos\gamma\cos\eta-\sin\gamma\sin\eta\cos\psi,
\label{eq:Theta}
\ee
where $\psi=(\Omega t+{\rm constant})$ is the rotation phase ($\psi=0$
when $\bmu$ lies in the $XZ$ plane).
We use the simplified formalism derived by \citet[][]{Beloborodov02a} to 
calculate the observed spectral flux from the area $dS$ of the hotspot, 
which takes the form:
\begin{eqnarray}
\label{eq:ObsFlux}
F_{\rm obs} = \left(1-{r_g\over R}\right)^{3/2}
I_{\nu}(\theta_{\rm em})\cos\theta_{\rm em} {dS\over D^2},
\end{eqnarray}
where $r_g=2GM/c^2$ is the Schwarzschild radius, $R$ is the NS radius, 
and $\theta_{\rm em}$ (the angle between the photon direction and the surface
normal at the emission point; $\mu=\cos\theta_{\rm em}$)
is related to $\Theta$ through:
\begin{eqnarray}
\label{eq:LBend}
\cos\theta_{\rm em}={r_g\over R}+\left(1-{r_g\over R}\right)\cos\Theta.
\end{eqnarray}
For $R>3 r_g$, the spectral flux calculated using eq.~(\ref{eq:ObsFlux})
differs from the exact treatment \citep[see][]{Pechenicketal83a} by
less than $1$\%.

\begin{figure}
\includegraphics[width=84mm]{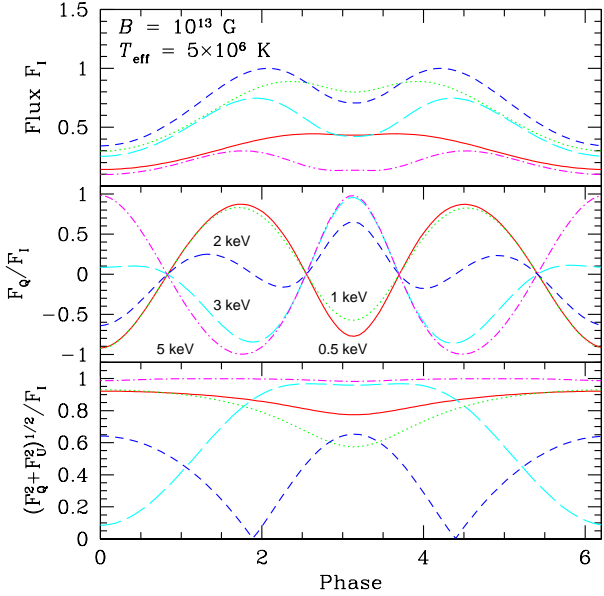}
\caption{Lightcurve and polarization as a function of rotation phase for a
NS hotspot with $B=10^{13}$ G, $T_{\rm eff}=5\times 10^6$ K.
The angle of the spin axis relative to the line of sight is
$\gamma=30\degr$, and the inclination of the magnetic axis relative to
the spin axis is $\eta=70\degr$.  Note that the sign of the $F_Q$
Stokes parameter is opposite for low and high energy photons; this
implies that the planes of polarization for low and high energy
photons are perpendicular.  This is a unique signature of vacuum
polarization for models with $B<B_l$.}
\label{fig:stokesb01}
\end{figure}

\begin{figure}
\includegraphics[width=84mm]{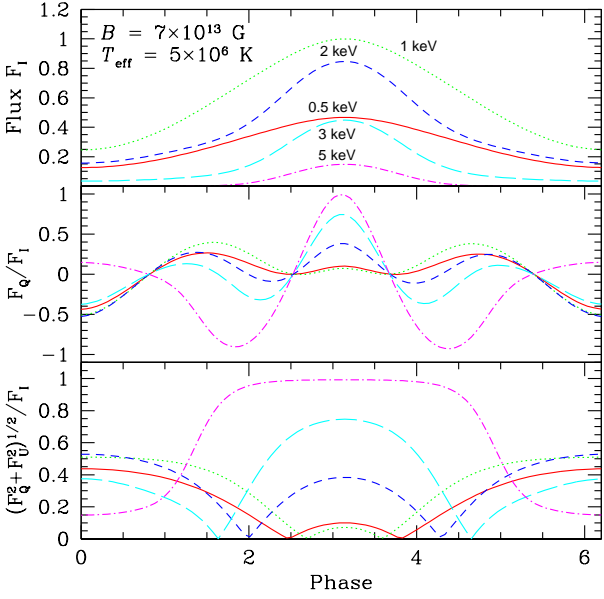}
\caption{Same as Fig.~\ref{fig:stokesb01} with $B=7\times 10^{13}$ G.
For this model, $B\sim B_l$, representing the transition between the
models shown in Figs.~\ref{fig:stokesb01} and \ref{fig:stokesb50}.}
\label{fig:stokesb07}
\end{figure}

\begin{figure}
\includegraphics[width=84mm]{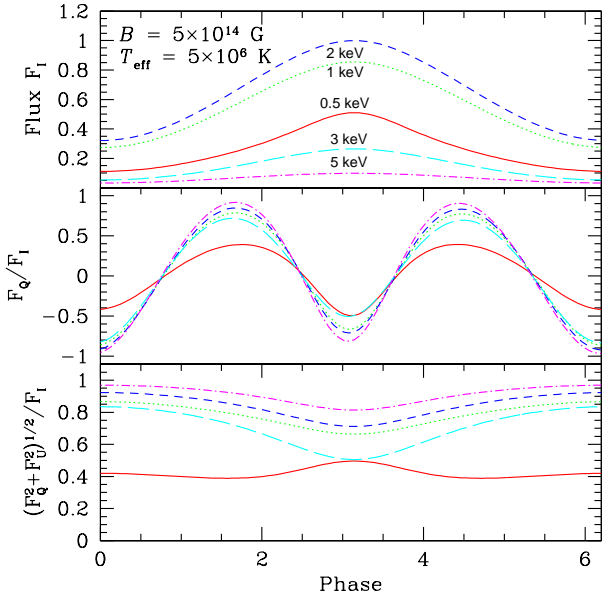}
\caption{Same as Fig.~\ref{fig:stokesb01} with $B=5\times 10^{14}$~G.
At this field strength, $B>B_l$, and the vacuum resonance lies between
the O and X-mode photospheres.  Thus, the sign of the $F_Q$ Stokes
parameter for low and high energy photons is the same.  Note that in
the top panel, the flux for the 0.5 keV case is multiplied by a factor
of 10 relative to the other curves.}
\label{fig:stokesb50}
\end{figure}

The top panels of Figs. \ref{fig:stokesb01}-\ref{fig:stokesb50} show
lightcurves for NS models with several magnetic field strengths, at a
range of energies $E=0.5-3$ keV, with geometry $\gamma=30\degr$,
$\eta=70\degr$.  

\subsection{Observed linear polarization signals}
\label{subsect:Polar}

The atmosphere models presented in \S\ref{sect:Method} and
\S\ref{sect:Results} give the specific intensities 
$I_\nu^{\rm X}(\theta_{\rm em})$ and $I_\nu^{\rm O}(\theta_{\rm em})$
of the two photon modes, 
emerging from the NS atmosphere
(outside the vacuum resonance layer). To determine the observed
polarization signals, it is important to consider propagation of the polarized 
radiation in the NS magnetosphere. In the X-ray band, the magnetospheric dielectric 
properties
are dominated by vacuum polarization (Heyl \& Shaviv 2002). 
Heyl et al.~(2003) (see also Heyl \& Shaviv 2002) evolved
the Stokes parameters along photon geodesics in the magnetosphere
and showed that the observed polarization is determined at the 
so-called ``polarization limiting radius,'' at a distance far from the NS surface.
Below we present a simple calculation of the propagation effect and observed
linear polarization (see also Lai \& Ho 2003b).

Consider a photon emitted at time $t_i$ from the hotspot, with
rotation phase $\psi_i=\Omega t_i$. The emission point has 
polar angle $\Theta_i$ (relative to the fixed $XYZ$ frame)  given by
eq.~(\ref{eq:Theta}) with $\psi=\psi_i$, and azimuthal angle $\varphi_i$
given by 
\be 
\tan\varphi_i={\sin\eta\,\sin\psi_i\over
\sin\eta\cos\gamma\cos\psi_i+\cos\eta \sin\gamma}.  
\label{eq:phii}\ee 
This is also the angle [$\varphi_i=\varphi_B(R)$]
between the projection of the magnetic axis in the $XY$ plane
and the $X$-axis. After the photon leaves the star,
it travels towards the observer, with a trajectory given by 
\be 
{\bf r}=(R\sin\Theta_i\cos\phi_i+\Delta_X)\,{\hat X}+(R\sin\Theta_i\sin\phi_i
+\Delta_Y)\,{\hat Y}+(R\cos\Theta_i+s+\Delta_Z)\,{\hat Z}, 
\ee 
where $s=c\Delta t=c(t-t_i)$, $\Delta_{X,Y,Z}$ are relativistic
corrections (which, as we will see shortly, are unimportant 
for the polarization signals), and $\hat X,~\hat Y,~\hat Z$ are unit vectors.
As the photon propagates in the magnetosphere, it will
``see'' a changing stellar magnetic field,
given by ${\bf B}=-\nabla (\bmu\cdot{\bf r}/r^3)$, where\footnote{We restrict the
propagation to the ``near zone'' of the star, i.e., $r<r_l=c/\Omega$.}
\be
\bmu=\mu\left[(\sin\eta\cos\gamma\cos\psi+\cos\eta\sin\gamma)\,{\hat X}
+\sin\eta\sin\psi \,{\hat Y}+(\cos\eta\cos\gamma-\sin\eta\sin\gamma\cos\psi) 
\,{\hat Z}\right],
\ee with $\psi=\Omega t=\psi_i+\Omega\Delta t=\psi_i+s/r_l$ (here
$r_l=c/\Omega$ is the radius of the light cyclinder).  The photon's
polarization state will evolve adiabatically following the varying
magnetic field that the photon experiences, up to the polarization
limiting radius $r_{\rm pl}$, beyond which the polarization is frozen.
Since we anticipate $r_{\rm pl}\gg R$, we only need to consider the region
far away from the NS: for $r\gg R$, the photon trajectory is
simply ${\bf r}\simeq s\,{\hat Z}$, and the magnetic field along the
photon path is ${\bf B}\simeq (2\mu_Z{\hat Z}-\mu_X{\hat X}-\mu_Y{\hat
Y})/r^3$, with $r\simeq s$. This magnetic field has a magnitude
\begin{eqnarray}
\label{eq:BMag}
B(s)={B_s\over 2}\left({R\over r}\right)^3\left[1+3
\left(\cos\eta\cos\gamma-\sin\gamma\sin\eta\cos\psi\right)^2\right]^{1/2},
\end{eqnarray}
where $B_s=2\mu/R^3$ is the magnitude of the (dipole) surface field at the 
magnetic pole.
The magnetic field is inclined at an angle
$\theta_{kB}$ to the line of sight, and makes an azimuthal angle
$\varphi_B$ in the $XY$ plane such that:
\begin{eqnarray}
\label{eq:cthkB}
\sin^2\theta_{kB}(s) & = &
{1-\left(\cos\eta\cos\gamma-\sin\gamma\sin\eta\cos\psi
\right)^2\over 1+3\left(\cos\eta\cos\gamma-\sin\gamma\sin\eta\cos\psi
\right)^2},\\
\label{eq:tphiB}
\tan\varphi_B(s)& = & {\sin\eta\sin\psi
\over \cos\eta\sin\gamma+\cos\gamma\sin\eta\cos\psi}.
\end{eqnarray}
Recall that in Eqs.~(\ref{eq:BMag})-(\ref{eq:tphiB}), $\psi=\psi_i+s/r_l
=\Omega t_i+s/r_l$ and $s$ is the affine parameter along the ray.

The wave equation for photon propagation in a magnetized medium takes
the form
\begin{eqnarray}
\label{eq:Maxwell}
\nabla\times (\bar{\mbox{\boldmath $\mu$}}\cdot\nabla\times{\bf E}) =
{\omega^2\over c^2}\mbox{\boldmath $\epsilon$}\cdot{\bf E},
\end{eqnarray}
where ${\bf E}$ is the electric field (not to be confused with the
photon energy), and {\boldmath $\epsilon$}, {\boldmath $\bar\mu$} are 
the dielectric and inverse permeability tensors, respectively.
In the magnetized vacuum of the NS magnetosphere, they are given by 
$\mbox{\boldmath $\epsilon$} =a{\bf I}+q{\bf\hat{B}}{\bf\hat{B}}$ and
$\mbox{\boldmath $\bar{\mu}$} =a{\bf I}+m{\bf\hat{B}}{\bf\hat{B}}$.
Solving eq.~(\ref{eq:Maxwell}) for EM waves with $E\propto e^{iks-i\omega t}$ 
yields the two modes (in the $XY$ basis)
\begin{eqnarray}
\label{eq:PolarModes}
{\bf e}_{\rm O}= (\cos\varphi_B,\sin\varphi_B),\ \ {\bf e}_{\rm X} =
(-\sin\varphi_B,\cos\varphi_B),
\end{eqnarray}
with indices of refraction $n_{\rm O} \simeq 1+(q/2)\sin^2\theta_{kB}$
and $n_{\rm X} \simeq 1-(m/2)\sin^2\theta_{kB}$. A general (transverse) EM
wave can be written as a superposition of the two modes:
\begin{eqnarray}
\label{eq:GenEM}
{\bf E} = {\cal A}_{\rm O} {\bf e_{\rm O}} + {\cal A}_{\rm X} {\bf e_{\rm X}}.
\end{eqnarray}
Following the steps of Lai \& Ho (2002) [see their eq.~(15)],
we derive the following equations for the evolution of the mode amplitudes:
\begin{eqnarray}
\label{eq:AmpEvol}
i \left(
\begin{array}{c}
{\cal A}_{\rm O}'\\
{\cal A}_{\rm X}'
\end{array}
\right)
& \simeq &
\left(
\begin{array}{cc}
-(\omega/c)\Delta n/2 & i\varphi_B'\\
-i\varphi_B' & (\omega/c)\Delta n/2
\end{array}
\right)
\left(
\begin{array}{c}
{\cal A}_{\rm O}\\
{\cal A}_{\rm X}
\end{array}
\right),
\end{eqnarray}
where the prime (') denotes a derivative with respect to $s$, and
$\Delta n=n_{\rm O}-n_{\rm X}={1\over 2}(q+m)\sin^2\theta_{kB}$.
The condition for adiabatic evolution of photon modes is 
\begin{eqnarray}
\label{eq:AdCond}
(\omega/c)\Delta n\gg 2\varphi_B'.
\end{eqnarray}
Near the star ($r\sim R$), $\varphi_B\sim 1/r$, and the adiabatic condition
is easily satisfied. Far from the star ($r\gg R$),
using eq.~(\ref{eq:BMag}), we write:
\be
\label{eq:Dn}
\Delta n = {\alpha\over 30\pi}\left({B\over B_Q}\right)^2
\sin^2\theta_{kB}
 = 9.94\times 10^{-9} B_{12}^2 \left({R\over r}\right)^6F_B,
\ee
where
\be
F_B = 1-\left(\cos\eta\cos\gamma-\sin\gamma\sin\eta\cos\psi\right)^2,
\ee
and $B_{12}=B_s/(10^{12}{\rm G})$. From eq.~(\ref{eq:tphiB}), we have
\be
\label{eq:Vphi'}
{d\varphi_B\over ds}  = {1\over r_l} F_{\varphi},
\ee
with
\be
F_{\varphi} = \left(\sin^2\eta\cos\gamma+\sin\eta\cos\eta\sin\gamma\cos\psi
\right)/F_B.
\ee
The polarization-limiting radius $r_{\rm pl}$ is set by 
the condition $\omega \Delta n/c = 2\varphi_B'$.
Substituting in eqs.~(\ref{eq:Dn}) and (\ref{eq:Vphi'}) we find
\footnote{Our expression for $r_{\rm pl}$ differs from that given in
Heyl \& Shaviv (2002) and Heyl et al.~(2003).}
\begin{eqnarray}
\label{eq:Rpl}
{r_{\rm pl}\over R}=32.6\left({E_1 B_{12}^2F_B\over
f_1\,F_{\varphi}}\right)^{1/6},
\end{eqnarray}
where $f_1$ is the spin frequency $\Omega/(2\pi)$ in Hz, 
and $F_B$, $F_{\varphi}$ are slowly varying functions of 
phase and are of order unity.  Note that the above analysis is valid
only if $r_{\rm pl}\lo r_l/2$, since beyond the light-cylinder radius
the magnetic field is no longer described by a static
dipole. Thus we require that
\be
{r_{\rm pl}\over r_l}\simeq 6.84\times 10^{-3}
\left({E_1 B_{12}^2F_B\over F_{\varphi}}\right)^{1/6}R_{10}\,f_1^{5/6}
\lo 0.5,
\label{eq:rplrl}\ee
where $R_{10}=R/(10~{\rm km})$.

Beyond $r_{\rm pl}$, the polarization state of the photon is ``frozen.''
Using eq.~(\ref{eq:PolarModes}), the observed Stokes parameters 
in the observer coordinate system ($XYZ$) are given by
\begin{eqnarray}
\label{eq:AdStokes1}
I & = & I_{\rm O}+I_{\rm X},\\
\label{eq:AdStokes2}
Q & \simeq & (I_{\rm O}-I_{\rm X}) \cos 2\varphi_B(r_{\rm pl}),\\
\label{eq:AdStokes3}
U & \simeq & 
(I_{\rm O}-I_{\rm X})\sin 2\varphi_B(r_{\rm pl}),
\end{eqnarray}
where $I_{\rm O}\propto I_\nu^{\rm O}(\theta_{\rm em})$ and $I_{\rm X}\propto 
I_\nu^{\rm X}(\theta_{\rm em})$ are the specific mode intensities emitted at 
the NS surface [calculated with our models described in \S\ref{sect:Method} 
and \S\ref{sect:Results}, and corrected for the general relativistic
effect; see eqs.~(\ref{eq:ObsFlux})-(\ref{eq:LBend})], 
and $\varphi_B(r_{\rm pl})$ is evaluated at $s\simeq r=r_{\rm pl}$.  
From eqs.~(\ref{eq:phii}) and (\ref{eq:tphiB}), with $\psi(r_{\rm pl})
=\psi_i+r_{\rm pl}/r_l$, we see that the effect of NS rotation is to shift the
polarization lightcurve by a phase $r_{\rm pl}/r_l$. For slow rotation
$r_{\rm pl}/r_l\ll 1$ [see eq.~(\ref{eq:rplrl})], this shift is small, and
$\varphi_B(r_{\rm pl})\simeq \varphi_i+\pi$. We calculate the observed spectral
fluxes $F_I=F,~F_Q,~F_U$ associated with the intensities $I,~Q,~U$
using the standard procedure described in \S \ref{subsect:LCurves}.

The middle and bottom panels of
Figs.~\ref{fig:stokesb01}-\ref{fig:stokesb50} show the phase evolution
of the Stokes parameter $F_Q$ and the degree of linear polarization, 
both normalized to the observed spectral flux.  Note that 
$Q$ is defined such that $Q=1$ corresponds to linear
polarization in the plane spanned by the line of sight $Z$
and the NS spin axis.  In Fig.~\ref{fig:stokesb01}, we consider
emission from a NS hotspot with $B=10^{13}$ G and $T_{\rm eff}=5\times
10^6$ K.  Note that the value of $F_Q$ for low energy photons ($E \lo 1$
keV) is of opposite sign to that of high energy photons ($E \go 3$ keV).
This implies that the planes of polarization for low and high energy
photons are perpendicular.  This is a unique signature of vacuum
polarization first identified by \citet[][]{LaiHo03b}, which occurs
for $B<B_l$ [see eq.~\ref{eq:maglim}] because
the vacuum resonance appears outside the O-mode
photosphere. Below the vacuum resonace layer ($\rho>\rho_V$), the X-mode 
flux dominates over the O-mode. For low energy photons, $E\lo E_{ad}$, 
mode conversion is inefficient, and the emergent flux is dominated
by the X-mode; for high energy photons $E\go E_{ad}$, mode
conversion is efficient, rotating the plane of polarization, and 
the emergent flux is dominated by the O-mode [see Fig.~2 of Lai \& Ho (2003b)].

\begin{figure}
\includegraphics[width=84mm]{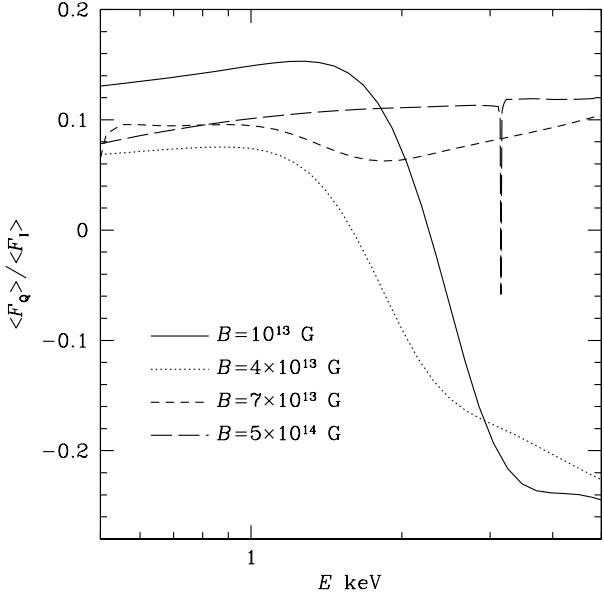}
\caption{Phase average of Stokes parameter $F_Q$ as a function of
photon energy for a rotating NS hotspot at magnetic field strengths
$B=10^{13}, 4\times 10^{13}, 7\times 10^{13}$ and $5\times 10^{14}$~G, with
$T_{\rm eff} = 5\times 10^6$~K.  Note that the sign of $\langle F_Q\rangle$ changes
between low and  high photon energies for the low-field cases, corresponding to
rotation of the plane of polarization.}
\label{fig:phaseave}
\end{figure}

Fig.~\ref{fig:stokesb50} shows the same result for the model with
$B=5\times 10^{14}$ G, $T_{\rm eff} = 5\times 10^6$ K.  In this case,
$B>B_l$, the vacuum resonance appears inside the O and X-mode
photospheres, and the emergent radiation is always dominated by the X-mode.
As expected, the planes of polarization for low and
high energy photons are aligned.  Fig.~\ref{fig:stokesb07} shows an
intermediate case, where $B\sim B_l$ --- to calculate the spectra and
polarization signals for such models, it is particularly important to 
incorporate partial mode conversion properly.
The distinct behavior between the low field and high field cases is illustrated by
Fig.~\ref{fig:phaseave}, which shows the phase-averaged $F_Q$ Stokes
parameter as a function of photon energy for several values of the magnetic field
strength.  The low-field cases show the characteristic rotation of
the plane of polarization between low-$E$ and high-$E$,
whereas the high-field cases do not.

Note that in the above analysis, the observed linear polarization
fraction $\Pi_L$ (the bottom panel of Figs.~\ref{fig:stokesb01}-\ref{fig:stokesb50})
is the same as the value just outside the emission region, 
$|\Pi_{em}|$, i.e., 
\be
\Pi_L={(Q^2+U^2)^{1/2}\over I}=|\Pi_{em}|,\qquad
{\rm with}\quad \Pi_{em}={I_\nu^O(\theta_{em})-I_\nu^X(\theta_{em})
\over I_\nu^O(\theta_{em})+I_\nu^X(\theta_{em})}.
\label{eq:piem}\ee
The polarized fluxes are simply 
\be
F_Q\simeq F_I\Pi_{em}\cos 2\varphi_B(r_{\rm pl}),\qquad 
F_U\simeq F_I\Pi_{em}\sin 2\varphi_B(r_{\rm pl}).
\ee
In \S\ref{subsect:CircPol} we shall see that for rapidly rotating NSs,
the observed $\Pi_L$ will be somewhat smaller than $|\Pi_{em}|$
because of the generation of circular polarization around $r_{\rm
pl}$.

\subsection{Circular polarization}
\label{subsect:CircPol}

Circular polarization of surface emission 
may be generated for sufficiently rapid rotating NSs 
due to the gradual wave mode coupling/decoupling around $r_{\rm pl}$.
While the linear polarization signals can be adequately described and 
calculated using the simple analysis given in \S \ref{subsect:Polar},
to calculate the circular polarization, quantitative
solutions of the evolution equations for the modes 
or the Stokes parameters in the magnetosphere are necessary.

Consider the mode evolution equations (\ref{eq:AmpEvol}). For given
initial values (e.g., mode amplitudes at $r_i\ll r_{\rm pl}$),
the solution of the equations depend on two parameters: $C$, defined
by $(\omega/c)\Delta n\approx C/r^6$ and $\varphi_B'$ (since $\psi$
varies by a rather small amount along the ray path, $F_B$ and
$F_\varphi$ are nearly constant).  Alternatively, since $r_{\rm pl}$
is determined by $C/r^6=2\varphi_B'$, the solution depends only on the
dimensionless parameter $\Gamma$, defined by
\be
\Gamma\equiv r_{\rm pl}\varphi_B'
={r_{\rm pl}\over r_l}\,F_\varphi
\simeq 6.84\times 10^{-3}
\left({E_1 B_{12}^2F_B F_{\varphi}^5}\right)^{1/6}R_{10}\,f_1^{5/6}.
\ee
Indeed,
if we define $x=r/r_{\rm pl}$, eq.~(\ref{eq:AmpEvol}) can be rewritten as
\begin{eqnarray}
\label{eq:AmpEvol2}
i{d\over dx} \left(\begin{array}{c}
{\cal A}_{\rm O}\\ {\cal A}_{\rm X}\end{array}\right)
& \simeq & \Gamma \left(\begin{array}{cc}
-x^{-6} & i\\ -i & x^{-6}\end{array}
\right)\left(\begin{array}{c}
{\cal A}_{\rm O}\\{\cal A}_{\rm X}\end{array}\right).
\end{eqnarray}
Note that while eq.~(\ref{eq:AmpEvol}) is valid for all radii, the
magnetostatic approximation for the field of a rotating dipole breaks
down beyond the light-cylinder radius. Since 
$\varphi_B'\sim 1/r_l$, eq.~(\ref{eq:AmpEvol2}) is valid
only for $\Gamma\sim r_{\rm pl}/r_l\lo 0.5$.

The Stokes parameters can be written in terms of the mode amplitudes as
\begin{eqnarray}
\label{eq:MagStokes}
I & = & |E_X|^2 + |E_Y|^2 = |{\cal A}_{\rm O}|^2+|{\cal A}_{\rm X}|^2,\\
Q & = & |E_X|^2 - |E_Y|^2 = \cos 2\varphi_B \,
(|{\cal A}_{\rm O}|^2-|{\cal A}_{\rm X}|^2)
-2\sin 2\varphi_B\, \Re e \left({\cal A}_{\rm O} {\cal A}_{\rm X}^\ast\right),\\
U & = & 2\,\Re e\left(E_X E_Y^*\right) = 
\sin 2\varphi_B\, (|{\cal A}_{\rm O}|^2-|{\cal A}_{\rm X}|^2) + 2\cos 2\varphi_B\, 
\Re e\left({\cal A}_{\rm O}{\cal A}_{\rm X}^*\right),\\
\label{eq:MagStokes4}
V & = & 2\,\Im m\left(E_XE_Y^*\right) = 
2\,\Im m\left({\cal A}_{\rm O}{\cal A}_{\rm X}^*\right).
\end{eqnarray}

\begin{figure}
\includegraphics[width=12cm]{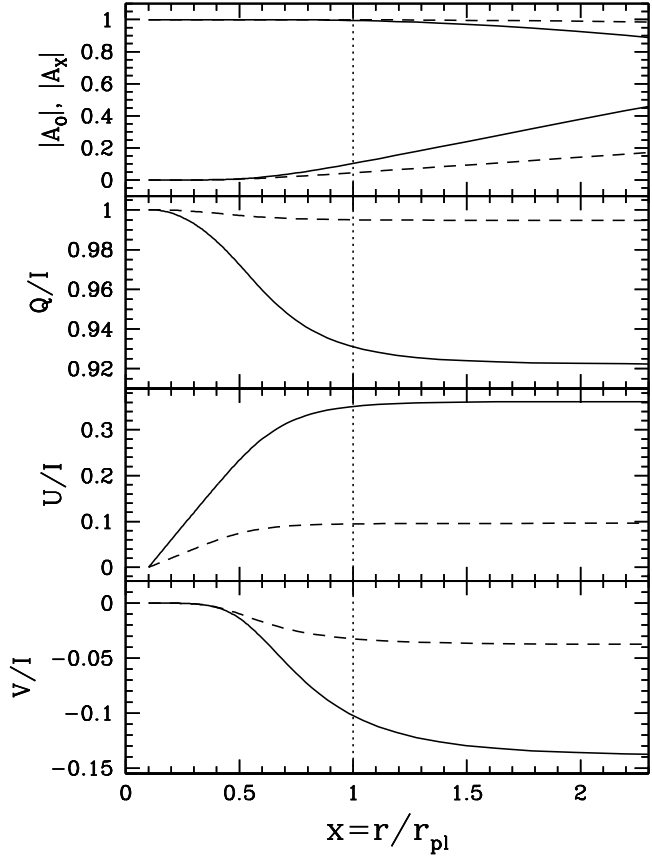}
\caption{Evolution of the radiation mode amplitudes (top panel) and
Stokes parameters (bottom three panels). The solid lines are for
$\Gamma=0.3$ and the dashed lines for $\Gamma=0.1$.  The polarization
limiting radius is shown as the vertical dotted line. The initial
values (at a small $x=x_i$) are ${\cal A}_{\rm O}=1,~{\cal A}_{\rm X}=0$,
$Q=I=1,~U=0$ and $V=0$.
At distances $x\lo 0.5$, the modes evolve adiabatically.  At $r\sim r_{\rm
pl}$ the modes begin to couple, generating appreciable circular
polarization.  At $x\go 2$, the values of the Stokes
parameters are ``frozen''.}
\label{fig:amp}
\end{figure}

Figure \ref{fig:amp} 
gives some examples of the results of numerical integration
of eq.~(\ref{eq:AmpEvol2}). We start the integration at 
radius $x_i=r_i/r_{\rm pl}$ such that the adiabatic condition is well
satisfied (we typically choose $x_i\lo 1/3$). In these examples,
the initial values are ${\cal A}_{\rm O}=1$, ${\cal A}_{\rm X}=0$.
After obtaining ${\cal A}_{\rm O}(x)$ and ${\cal A}_{\rm X}(x)$, we calculate
the Stokes parameters using eqs.~(\ref{eq:MagStokes})-(\ref{eq:MagStokes4})
with $\varphi_B=\varphi_B'(r-r_i)=\Gamma (x-x_i)$
(adding a constant to $\varphi_B$ will affect $Q$ and $U$,
but not $V$). We see that for $x\lo 1/2$,
the photon modes evolve adiabatically, and thus $Q\propto
\cos 2\varphi_B$, $U\propto \sin 2\varphi_B$ [see eqs.~
(\ref{eq:AdStokes2})-(\ref{eq:AdStokes3})] and $V\simeq 0$.  Around
$x=1$ ($r=r_{\rm pl}$), the modes couple and circular polarization is generated.
For $x\go 2$, the values of the Stokes parameters are
``frozen'' and no longer evolve. NSs with large $\Gamma$
(corresponding to rapid rotations; e.g., $f=40$~Hz, $B=10^{13}$~G, $E=1$~keV
yields $\Gamma\sim 0.3$) can generate
appreciable circular polarization, with $V(r\rightarrow\infty)/I
\approx -14\%$. As the NS spin frequency decreases, the resulting
$|V/I|$ decreases.

Alternatively, using the relations 
(\ref{eq:MagStokes})-(\ref{eq:MagStokes4}) and the mode 
evolution equation (\ref{eq:AmpEvol}), we can derive
evolution equations for the Stokes parameters:
\begin{eqnarray}
\label{eq:STE1}
I' & = & 0,\\
\label{eq:STE2}
Q' & = & (\omega\Delta n/c) \,V\, \sin 2\varphi_B,\\
U' & = & -(\omega\Delta n/c) \,V\, \cos 2\varphi_B,\\
\label{eq:STE4}
V' & = & -(\omega\Delta n/c) \,Q\,\sin 2\varphi_B + (\omega\Delta n/c)\,
U\,\cos 2\varphi_B.
\end{eqnarray}
Since the vacuum contribution to the dielectric tensor includes no
dissipation, $I'=0$ as expected.
Equations (\ref{eq:STE1})-(\ref{eq:STE4}) can be evolved
numerically to calculate the observed Stokes parameters.
Again we start the integration at a
radius $r_i<r_{\rm pl}$ such that the adiabatic condition is well
satisfied. Since the circular polarization does not depend on the 
orientation of the $XY$ axes, we set the initial conditions at $r_i$ 
by rotating the coordinate system azimuthally so that $I=1$, $U=0$ (this also
corresponds to choosing $\psi=0$ or $\varphi_B=0$ at $r=r_i$),
and $Q_i=\Pi_{em}\le 1$, where $\Pi_{em}$ is the linear polarization
fraction just outside the atmosphere [see eq.~(\ref{eq:piem})].  Since
the radiation emerging from the NS surface is linearly polarized, we
set $V(r_i)=0$. Equations (\ref{eq:STE1})-(\ref{eq:STE4}) are then
integrated to a distance beyond $r_{\rm pl}$. 

For a given initial value of the linear polarization fraction $\Pi_{em}$
at a small $r_i\ll r_{\rm pl}$,
the solution of eqs.~(\ref{eq:STE2})-(\ref{eq:STE4})
depends only on the dimensionless parameter
$\Gamma\equiv r_{\rm pl}\varphi_B'$. Again, if we define $x=r/r_{\rm pl}$,
eqs.~(\ref{eq:STE2})-(\ref{eq:STE4}) can be rewritten as
\begin{eqnarray}
dQ/dx & = & {2\Gamma\over x^6}\,V\, \sin 2\varphi_B,\label{eq:dq}\\
dU/dx & = & -{2\Gamma\over x^6}\,V\, \cos 2\varphi_B,\label{eq:du}\\
dV/dx & = & -{2\Gamma\over x^6}\left(Q\,\sin 2\varphi_B - 
U\,\cos 2\varphi_B\right),\label{eq:dv}
\end{eqnarray}
with $\varphi_B=\Gamma (x-x_i)$. We are interested in 
the value of $V$ at $x=x_f\gg 1$. Figure \ref{fig:amp} shows some
examples of the integration of Eqs.~(\ref{eq:dq})-(\ref{eq:dv}). 
Not surprisingly,
the results are in exact agreement with those obtained
using the mode evolution equations.
 
\begin{figure}
\includegraphics[width=15cm]{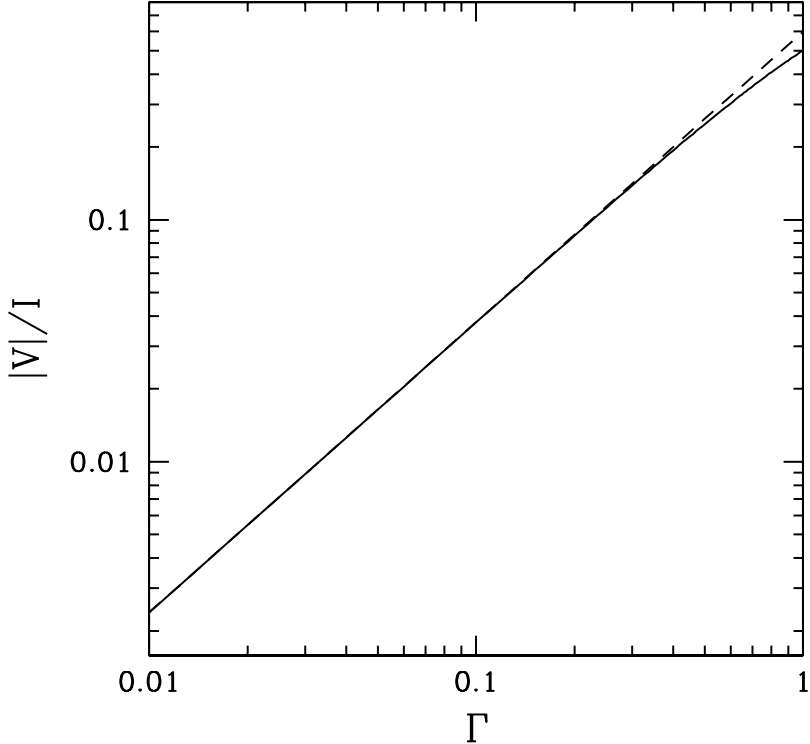}
\vskip -2cm
\caption{The magnitude of the observed circular polarization fraction 
$|V|/I$ as a function of $\Gamma$. The linear polarization fraction
($\Pi_{em}$) just outside the atmosphere is assumed to be $100\%$.
The dashed line depicts the fitting formula, eq.~(\ref{eq:Vobs}),
which agrees with the numerical solution to within $1\%$ for $\Gamma<0.4$.
Note that solutions with $\Gamma\go 0.5$ are incorrect
since the magnetic field around the polarization limiting radius is
no longer described by the near-zone field of a rotating dipole 
as adopted in our calculations.}
\label{fig:amp2}
\end{figure}

We have calculated the circular polarizations produced by 
rotating NSs with different values of $\Gamma$. Our 
numerical results [see Fig.~(\ref{fig:amp2})]
show that the generated circular polarization is given by the 
expression
\begin{eqnarray}
\label{eq:Vobs}
V/I\approx -0.60\,\Pi_{em}\, \mbox{sign}(\varphi_B')
\left|r_{\rm pl}\,\varphi_B'\right|^{6/5} 
= -1.5\times 10^{-3}\,\Pi_{em}\,\left(E_1\,B_{12}^2\,F_B\right)^{1/5}
\,f_1\,F_{\varphi}.
\end{eqnarray}
This expression is accurate to within one percent
in the regime $\Gamma\lo 0.4$ [see Fig.~(\ref{fig:amp2})].
Recall that $F_B\sim F_\varphi\sim 1$, so eq.~(\ref{eq:Vobs})
provides a quick estimate for the magnitude of the circular polarization
of NS surface emission.

The above analysis shows that substantial circular polarization can be
generated only for NSs with sufficiently rapid rotation and magnetic
field strength.  Given the photon energy, NS spin frequency, and
dipole magnetic field strength, the degree of circular polarization
from a NS can be calculated using eq.~(\ref{eq:Vobs}).

\begin{figure}
\includegraphics[width=84mm]{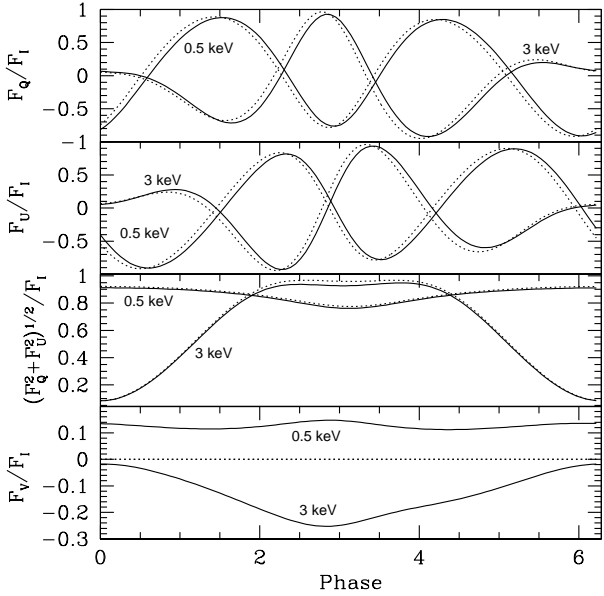}
\caption{Observed radiation Stokes parameters for a NS with
$B=10^{13}$~G, $T_{\rm eff}=5\times 10^6$~K, $f=50$~Hz,
$\gamma=30\degr$, and $\eta=70\degr$, for photon energies $E=0.5, 3$
keV.  The solid curves show the results of numerical integration of the 
transfer equations for the Stokes parameters, while the dotted curves are
calculated using the approximate method of \S~\ref{subsect:Polar}.
}\label{fig:fast}
\end{figure}

Fig.~\ref{fig:fast} shows the phase-resolved observed radiation Stokes
parameters for a NS with $B=10^{13}$~G and $T_{\rm eff}=5\times 10^6$~K,
rotating at $f=50$ Hz, with magnetic field and spin geometry
$\gamma=30\degr$, $\eta=70\degr$ (this is the case shown in
Fig.~\ref{fig:stokesb01}).  
The solid curves are numerical solutions to
the Stokes parameter equations of transfer in the NS magnetosphere,
while the dotted curves are calculated using the method described in
\S~\ref{subsect:Polar}.  The latter method assumes $F_V=0$, but yields
results for $F_Q$ and $F_U$ that are quite close to those of
the numerical integrations. For a rapidly rotating
NS, substantial circular polarization is generated, with
$|F_V/F_I|$ reaching $0.2-0.3$ in the hard X-ray band
(see also Fig.~\ref{fig:amp2}).

Finally, we note that although the specific results presented
in this section refer to emission from a hot polar cap on the NS,
we expect many of our key results (e.g., 
rotation of the planes of
linear polarization between $E\lo 1$~keV and at $E\go 4$~keV due to 
vacuum polarization 
for $B\lo 7\times 10^{13}$~G) to be valid in more complicated models
(when several hotspots or the whole stellar surface contribute to the
X-ray emission). This is because the polarization-limiting radius (due to
vacuum polarization in the magnetosphere) lies far away from the star
[see eq.~\ref{eq:Rpl}] where rays originating from different patches of 
the NS experience the same dipole field (Heyl et al.~2003).
Our results therefore demonstrate the unique potential of X-ray
polarimetry in probing the physics under extreme conditions 
(strong gravity and magnetic fields) and the nature of various forms 
of NS.


\section{Discussion}
\label{sect:Discussion}

We have presented a new method for incorporating partial conversion of 
photon modes due to vacuum polarization into fully-ionized, self-consistent 
atmosphere models 
of magnetized NSs.
This method takes into account the 
nontrivial probability of photon 
mode conversion at the vacuum resonance.
While recent works have clearly identified the 
important effects of the vacuum resonance and related mode conversion
in determining the atmosphere radiation spectrum and polarization 
(Lai \& Ho 2003a), so far the implementation of these effects
in self-consistent atmosphere models, for technical reasons, 
has only considered two extreme limits: 
complete mode conversion and no mode conversion (Ho \& Lai 2003,2004).
With a direct, semi-explict Runga-Kutta integration of the radiative 
transfer equations for the photon modes 
(as opposed to the forward-backward substitution 
procedure of the Feautrier method) and the use of an accurate mode
conversion formula for each photon, our new atmosphere code
displays excellent stability with respect to grid resolution. Moreover, 
integration of the full transfer equations for the radiation Stokes 
parameters shows that our treatment of partial conversion is accurate. 
As expected, the partial conversion solution is intermediate between 
the extreme cases of complete and no conversion considered previously.
An accurate treatment of vacuum polarization is a critical step toward 
interpreting the spectra and 
predicting the polarization signals of magnetic NSs.

With our new atmosphere code, we have constructed a large number of
atmosphere models for various magnetic field strengths, ranging from
$10^{13}$~G to $5\times 10^{14}$~G, for both H and He compositions.
In agreement with previous, approximate calculations (Ho \& Lai
2003,2004), we find that for $B\go 7\times 10^{13}$~G the vacuum
resonance affects the atmosphere spectra (i.e., the hard spectral tail
due to nongrey opacities tends to be suppressed and the spectral line
%
%
EW is reduced), and the effects become more significant as the magnetic 
field strength is increased. 
For $B\lo 7\times 10^{13}$~G, the effect of the vacuum resonance 
on the spectra is smaller and becomes negligible for 
$B\lo 4\times 10^{13}$~G. However, even for such ``low''
field strengths, vacuum resonace has a significant effect on the observed
X-ray polarizations (see Lai \& Ho 2003b). 
Our new calculations presented in this paper 
are particularly important for the ``intermediate''
field regime ($4\times 10^{13}\lo B\lo 2\times 10^{14}$~G), where previous
approximate treatments are inadequate.
For the first time, we are able to accurately determine the $B$-dependence 
of the structure, spectra and polarization signals of ionized NS 
atmospheres.

Since the most time-consuming and difficult part of atmosphere modeling
involves finding the atmosphere temperature profile that satisfies 
the condition of radiative equilibrium, for the convenience of the
astrophysics community, we have presented fitting formulae for the 
temperature profiles of various atmosphere models (see \S\ref{subsect:AtmoStruct}). 
With these analytic expressions, it is relatively
straightforward (using the procedure outlined in \S\ref{subsubsect:ModeSol}) to calculate
various properties of the emergent radiation. These analytic profiles
will also be useful for comparison with future theoretical atmosphere 
models. 

We note that the models presented in this paper have several limitations: 
(1) The models assume that the magnetic field lies along the surface 
normal. While a more general magnetic field inclination 
does not change the main results of our paper (e.g. the effect 
of vacuum resonance and the dependence of the atmosphere spectra on 
$B$), to confront observations (see below), 
synthetic spectra must be constructed 
using realistic magnetic field and surface temperature distributions, 
adding up contributions over the entire NS surface.  
Such calculations are necessarily model-dependent, but they are needed for 
%
%
proper interpretation of observations.\footnote{Synthetic spectra 
from NSs with realistic temperature distributions may provide a solution 
to the problem of the XDINS optical excess discussed in \S\ref{sect:Introduction} 
\citep[see, e.g.,][]{Ponsetal02a,Burwitzetal03a}.}

(2) At high density, the radiative transfer equation breaks down, due 
to the dense plasma effect.
At large optical depth, the photon polarization develops a non-negligible
longitudinal component and the index of refraction deviates
significantly from unity.  This occurs when the plasma frequency of
the medium exceeds the photon frequency. 
To date, no detailed studies of the
transfer of radiation in dense plasmas have been performed, though it
has been treated in an ad-hoc way by \citet[][]{Hoetal03a}.
Nevertheless, this effect is important for treating thermal radation
in the optical band, and for magnetars can affect the emission
spectrum in the soft X-ray ($\lo 1$~keV).
(3) The assumption of fully ionized atmospheres may not be valid for cool
NSs (such as the dim isolated NSs) or 
even the higher temperature AXPs and SGRs. Nevertheless, we expect
features due to bound-bound and bound-free transitions of neutral
species to be suppressed in the same manner as the ion cyclotron feature 
(see Ho et al.~2003; Potekhin et al.~2004,2005 for recent works
on partially ionized magnetic atmosphere models).

\subsection{Implications for observations of isolated neutron stars}

As mentioned in 
\S 1, recent observations by 
\it{Chandra}\rm\ and \it{XMM-Newton}\rm\ have shown that 
the quiescent thermal spectra from AXPs and SGRs have 
no observable absorption features, 
such as the ion cyclotron line at $E_{Bi}=0.63(Z/A)B_{14}$~keV.
As first pointed out by Ho \& Lai (2003), and confirmed by 
our more accurate calculations presented here, 
the inclusion of vacuum polarization effects provides a natural 
explanation for the non-detection: at $B=5\times 10^{14}$~G, 
%
%
the vacuum-suppressed EW of the H or He cyclotron line is
smaller than the current detector resolution.  
We expect that in the magnetar field regime,
bound-bound and bound-free features will be similarly suppressed
(see Ho et al.~2003; Potekhin et al.~2004).

Prominent absorption lines (at $0.7$~keV and 1.4~keV) have been
detected from the source 1E1207.4-520, a young neutron star 
($T\simeq 2\times 10^6$~K) associated with 
a supernova remnant 
%
%
\citep[][]{Sanwaletal02a,Bignamietal03a,DeLucaetal04a,Morietal05a}
Two viable (but tentative) identifications of these 
features are: (1) Ion cyclotron and atomic transitions of light-element 
(most likely He) atmospheres at $B\go 10^{14}$~G (Pavlov \& Bezchastnov
2005); (2) Atomic transitions of C or O atmospheres with $B\lo 10^{12}$~G
(Mori et al.~2005). Based on our general result of line suppression in
the magnetar field regime, we suggest that the first interpretation is 
unlikely to be correct, although a quantitative calculation of the 
%
%
atomic line strengths is needed to draw a firm conclusion.\footnote{The energy 
spacing of the lines at 0.7, 1.4, 2.1, and 2.8~keV is strongly suggestive 
of cyclotron harmonics.
However, the analysis 
of \citet[][]{Morietal05a} indicates that, in all likelihood, the features at 
2.1 and 2.8~keV are not real.  Furthermore, a pure cyclotron line interpretation 
of these features is problematic because the strengths of the harmonics (either 
for electrons at $10^{10}$ G or protons at $10^{14}$~G) are expected to be 
negligible.}

Absorption features have also been detected from three nearby, dim
isolated NSs: RX J1308.6+2127, RX J1605.3+3249, and RX J0720.4-3125.
While all three sources have similar effective temperatures 
($T_{\rm eff}\sim 10^6$ K), their observed features occur at 
different energies and have varying equivalent widths:
$E\approx 0.2-0.3$ keV with EW$\approx 150$ eV for RX J1308.6+2127 
\citep[][]{Haberletal03a},
$E\approx 0.27$ keV with EW$\approx 40$ eV for RX J0720.4-3125 
\citep[][]{Haberletal04a}, and
$E\approx 0.45$ keV with EW$\approx 80$ eV for RX J1605.3+3249 
\citep[][]{vanKerkwijketal04a}.  
With a single line, it is difficult to conclusively determine the 
true atmosphere composition. 
One possibility is that these features are 
associated with proton cyclotron resonance (with possible
blending from atomic transitions) in a hydrogen atmosphere (Ho \& Lai 2004).
For RX J1308.6+2127, the inferred magnetic field is 
$3-5\times 10^{13}$~G,
for which line suppression by vacuum resonance is ineffective.
The broad width of the feature is consistent with that calculated by 
our hydrogen atmosphere models
(see Fig.~\ref{fig:Specb4t13}).  
RX J1605.3+3249 is also consistent with this picture: its feature 
corresponds to $B\sim 7\times 10^{13}$ G, and
partial suppression of the line may 
account for its lower (by a factor of $\sim 2$) EW
(see Fig.~\ref{fig:Specb7t13}).  
The situation for RX J0720.4-3125 is more complicated: the spectrum
(including the line width) varies as a function of the rotation 
phase (Haberl et al.~2004a)
and over a long timescale (a few years) (see Haberl et al.~2006).
If its absorption feature is a proton cyclotron line, then the 
inferred magnetic field 
is too low for vacuum polarization effects to alter the line strength. 
Its small EW ($40$~eV) could arise if the line-emitting region 
(where $B\la 10^{14}$ G) is a small fraction of the NS surface; 
most of the surface would have $B\ga 10^{14}$~G,
requiring a highly non-dipolar surface field.
Alternatively, if the atmosphere of RX J0720.4-3125 is composed of helium, 
the required field strength is $B\sim 9\times 10^{13}$~G, 
strong enough for the vacuum effects to reduce the line width.

Several recent papers have identified similar absorption features in 
other dim isolated NSs, though these observations may require independent 
confirmation and/or better statistics.  \citet[][]{Haberletal04b} report 
spectral features for RX J0806.4-4123 ($E\approx 0.4-0.46$ keV; 
EW$\approx 33-56$) and RX J0420.0-5022 ($E\approx 0.3$ keV; 
EW$\approx 45$), while \citet[][]{Zaneetal05a} report a spectral feature 
for RX J2143.7+0654 ($E\approx 0.75$ keV; EW$\approx 27$).
These features are similar to those described in more 
detail above, and suffer from the same difficulties of identification.  
Needless to say, since the magnetic field strengths of these NSs
likely lie in the range $5\times 10^{13}-10^{14}$~G, for which
accurate treatment of the vacuum resonance effect is crucial,
the atmosphere models developed in this paper will be particularly
useful, especially when combined with detailed modeling of (phase-dependent)
synthetic spectra and (energy-dependent) lightcurves.


\subsection{Implications for future works}

It is clear that further theoretical modeling of NS surface
emission is needed to confront observations. Our discussion 
above (\S 6.1) also suggests that even accurate theoretical
models and high-quality data may still be inadequate to break some of the
degeneracy (e.g., magnetic field strength and geomery, atmosphere
composition, surface temperature distribution)
inherent in the problem. In this regard, X-ray polarimetry is
highly desirable. Our calculations in \S\ref{sect:LCurvesPolar} show that polarization 
signals are complementary to X-ray spectra. For example,
the polarization signals of magnetars and NSs with ``ordinary'' field
strengths are qualitatively different. It is possible for 
a NS with a ``boring'' spectrum and lightcurve to generate
an interesting polarization signature!

\section*{Acknowledgments}
We thank Wynn Ho and Alexander Potekhin for several useful discussions.
%
%
We also thank Marten van Kerkwijk for pointing out an initial 
problem with our temperature profile fits, and for making 
several useful suggestions.
This work has been supported in part by NSF grant AST 0307252,
NASA grant NAG 5-12034 and {\it Chandra} grant TM6-7004X
(Smithsonian Astrophysical Observatory).  
M.V.A. was also supported in part by a fellowship from the NASA/New York 
Space Grant Consortium.


\end{document}